\documentclass[a4paper]{article}
\usepackage[utf8]{inputenc}
\usepackage{authblk}

\title{Learning efficient backprojections across cortical hierarchies in real time}

\author[1]{Kevin Max}
\author[1]{Laura Kriener}
\author[2]{Garibaldi Pineda García}
\author[2]{\\Thomas Nowotny}
\author[1]{{Ismael Jaras}}
\author[1]{Walter Senn}
\author[1]{Mihai A.~Petrovici}
\affil[1]{\small Department of Physiology, University of Bern, Switzerland}
\affil[2]{School of Engineering and Informatics, University of Sussex, Brighton, United Kingdom}
\date{\small\today}

\usepackage[final]{changes}
\definechangesauthor[color=cyan]{JJ}

\usepackage{lineno}

\usepackage[
    natbib=true,
    backend=bibtex,
    citestyle=numeric-comp,
    sorting=none,
    giveninits=true,
    uniquename=init
]{biblatex}
\usepackage{graphicx}
\usepackage{xcolor}
\usepackage{amsmath}
\usepackage{amsthm}
\usepackage{amssymb}
\usepackage{adjustbox}
\usepackage[ruled, vlined, linesnumbered]{algorithm2e}
\usepackage{amsthm,amsmath,amsfonts,amssymb,amscd}   %
\usepackage{bookmark}
\usepackage{booktabs}       %
\usepackage{bbm}
\usepackage{bm}
\usepackage[font=footnotesize,labelfont=bf]{caption}
\usepackage[capitalize, nameinlink]{cleveref}
    \creflabelformat{equation}{#2#1#3}
\usepackage{cuted}
\usepackage{enumitem}
\usepackage{geometry}
\usepackage[]{glossaries-extra}
\usepackage{graphicx}
\usepackage{listings}                                %
\usepackage{mathtools}
\usepackage{microtype}      %
\usepackage{multicol}
\usepackage{multirow}
\usepackage{nicefrac}       %
\usepackage{placeins}                                %
\usepackage{relsize}
\usepackage{setspace}
\usepackage{sidecap}
\usepackage[binary-units]{siunitx}
\usepackage{threeparttable}
\usepackage{tikz}
\usepackage{url}            %
\usepackage{wrapfig}
\usepackage{xcolor}         %
\usepackage{xspace}
\usepackage{mathtools}
\usepackage{subcaption}
\usepackage{changes}
\usepackage{scalerel,stackengine}

\newcommand{\api}{\text{api}}

\newcommand{\bas}{\text{bas}}
\newcommand{\brevex}[2]{{\breve{#1}}^{#2}}

\newcommand{\Cm}{C_\text{m}}
\newcommand{\den}{\text{den}}
\newcommand{\El}{E_\text{l}}

\newcommand{\gl}{g_\text{l}}
\newcommand{\gbas}{g^\text{bas}}
\newcommand{\gapi}{g^\text{api}}
\newcommand{\gden}{g^\text{den}}

\newcommand{\lambdaP}{\lambda^\text{P}}

\newcommand{\taueff}{\tau^\text{eff}}
\newcommand{\taueffP}{\tau^\text{eff,P}}
\newcommand{\taueffI}{\tau^\text{eff,I}}
\newcommand{\tauhp}{\tau_\text{hp}}
\newcommand{\taulo}{\tau_\text{lo}}

\newcommand{\tauxi}{\tau_{\xi}}

\newcommand{\Tpres}{T_\text{pres}}

\newcommand{\ubreve}{\breve{u}}

\newcommand{\ubreveI}{\brevex{u}{\text{I}}}
\newcommand{\ubreveP}{\brevex{u}{\text{P}}}
\newcommand{\ubrevePn}{\brevex{u}{\text{P,0}}}
\newcommand{\ubreven}{\brevex{u}{\text{0}}}
\newcommand{\vbashat}{{\hat{v}}^\text{bas}}

\newcommand{\vapi}{v^\text{api}}
\newcommand{\rI}{r^\text{I}}
\newcommand{\rP}{r^\text{P}}
\newcommand{\rPI}{r^\text{P/I}}
\newcommand{\WPP}{W^\text{PP}}
\newcommand{\WIP}{W^\text{IP}}
\newcommand{\BPP}{B^\text{PP}}
\newcommand{\BPI}{B^\text{PI}}

\newcommand{\E}{\mathbb{E}}

\newcommand{\tP}{\mathrm{P}}

\newcommand\equalhat{\mathrel{\stackon[1.5pt]{=}{\stretchto{%
				\scalerel*[\widthof{=}]{\wedge}{\rule{1ex}{3ex}}}{0.5ex}}}}

\newcommand{\id}{\mathbf{1}}

\begin{document}

\maketitle

\begin{abstract}
 
    Models of sensory processing and learning in the cortex need to efficiently assign credit to synapses in all areas.
    In deep learning, a known solution is error backpropagation, which however requires biologically implausible weight transport from feed-forward to feedback paths.
    We introduce Phaseless Alignment Learning (PAL), a bio-plausible method to learn efficient feedback weights in layered cortical hierarchies.
    This is achieved by exploiting the noise naturally found in biophysical systems as an additional carrier of information.
    In our dynamical system, all weights are learned simultaneously with always-on plasticity and using only information locally available to the synapses.
    Our method is completely phase-free (no forward and backward passes or phased learning) and allows for efficient error propagation across multi-layer cortical hierarchies, while maintaining biologically plausible signal transport and learning.
    Our method is applicable to a wide class of models and improves on previously known biologically plausible ways of credit assignment: compared to random synaptic feedback, it can solve complex tasks with fewer neurons and learn more useful latent representations.
    We demonstrate this on various classification tasks using a cortical microcircuit model with prospective coding.
\end{abstract}

\section{Introduction}

\replaced
{The two fields of deep learning and neuroscience remain at vastly different levels in the description of their respective subjects.
}
{
Deep learning has originally been inspired by neuroscience, being influenced by the description of the visual cortex in particular.
Nonetheless, these two fields remain at vastly different levels in the description of their respective subjects.
}
While deep learning has made great leaps in terms of applicability and real-world usage in the past decade, the study of biological neural systems has revealed a plethora of different brain areas, connection types, cell types, and neuron as well as system states.
Currently, no clear organization scheme of computations and information transfer in the brain is known,
and the question of how artificial neural networks (ANNs) are related to models of the cortex remains an active field of research~\cite{Yamins2016-dw,richards2019deep,Lillicrap2020-ev}.

However, progress is being made in bridging the gap between these two fields~\cite{roelfsema_attention-gated_2005,costa_cortical_2017,scellier2017equilibrium,10.1162/NECO_a_00949,sacramento2018dendritic,haider2021latent,lillicrap2014random,Yamins2016-dw,payeur2021burst,10.3389/fncom.2016.00094,richards2019deep}.
In particular, important similarities between cortical and artificial information processing have been highlighted:
as in the cortex~\cite{HAAK201873}, most ANN architectures process information hierarchically.
\deleted{Additionally, external stimuli generate activity in functional units (neurons), which utilize bottom-up and top-down information}%
Neural activity is modulated through learning, i.e.~long-term adaptation of synaptic weights.
However, it is currently unclear how weights are adapted across the cortex in order to competently solve a task.
This is commonly referred to as the credit assignment problem, where neuroscience may learn from deep learning~\cite{friedrich_spatio-temporal_2011,richards2019deep}.

In the case of ANNs, \replaced{a}{an efficient} solution to this problem is known: currently, error backpropagation~(BP)~\cite{rumelhart1986learning,lecun1988theoretical} is the gold standard\deleted{ for learning in artificial networks}.
However, BP has several biologically implausible requirements.
ANNs trained with BP operate in distinct\added{, alternating} forward\added{ (inference)} and backward\added{ (learning)} phases\deleted{, where inference and learning alternate}.
Between phases, network activities need to be buffered -- i.e., information is processed non-locally in time.
Furthermore, error propagation occurs through weights which need to be mirrored at synapses in different layers~(weight transport problem).

In order to explain credit assignment for analog, physical computing (in the cortex or on neuromorphic hardware), physically plausible architectures and algorithms are therefore needed.
We assume such dynamical systems to operate in continuous time.
\deleted{They may minimize the difference between network output and target (`cost') by performing (approximate) gradient descent.}
Ideally, \deleted{such} physical systems are able to learn from useful instructive signals at all times, using only information which is locally available in space and time.
Crucially, all physical systems have inherent sources of noise -- in the form of stochastic activity, noisy parameters or intrinsic fluctuations of electrical and chemical signals.
The theory we propose makes use of neuronal noise as an additional carrier of information, instead of treating it as a nuisance\deleted{ parameter}.

For efficient credit assignment\deleted{ as in ANNs}, errors need to be propagated from higher to lower areas in the \deleted{cortical }hierarchy.
\deleted{With vanilla BP being excluded due to biological implausibility of weight transport, the question of how such error propagation occurs remains open.}
\deleted{Several methods have been proposed where feedback connections are assumed to be fixed or are learned.}
Broadly, \replaced{approaches}{these} can be categorized into methods with fixed feedback connections (feedback alignment, FA~\cite{lillicrap2014random,nokland2016direct}), bio-plausible approximations to BP~\cite{374486,akrout2019deep, lansdell2019learning,ernoult2022towards}, or alternative cost minimization schemes~\cite{bengio2014auto,lee2015difference,meulemans2020theoretical,meulemans2021credit}.
In this work, we introduce a method in the second category, related to the top-down weight alignment method of Ref.~\cite{ernoult2022towards}, which itself is based on previous insights on cost minimization in difference target propagation~\cite{meulemans2020theoretical}.
\deleted{The aim of our algorithm is to propagate \added{a} BP-like error, and to perform gradient descent on a cost function.}

\added{Our theory improves on existing literature in several ways:}\replaced{ W}{ is that w}e propose a fully dynamical system with efficient always-on plasticity\replaced{.}{:}
\replaced{T}{t}he neuronal and weight \added{continuous-time }dynamics model properties of physical substrates\deleted{, while learning is completely phaseless,} and plasticity is enabled for all synapses and at all times.
In agreement with biological plausibility, our method allows for efficient learning without requiring wake-sleep phases or other forms of phased plasticity implemented in many other models of learning in the cortex~\cite{6796552,ackley_learning_1987,bengio2015early,sacramento2018dendritic,guerguiev2017towards,scellier2017equilibrium,mesnard2019ghost,10.1162/089976603762552988,Song2022.05.17.492325}.
Our method is based on modeling of biologically plausible signal transport\replaced{, and the}{ in the form of rate-coding. The} learning mechanism incorporates bio- and hardware-plausible \deleted{components and }computations;
all dynamics and plasticity rules are fully local in time and space.
\deleted{Our model also makes full use of a recently proposed prospective coding mechanism~\cite{haider2021latent}.
This ensures fast propagation of information through layered networks, leading to quick convergence of useful top-down projections when using our method.}

\section{Results}

\subsection{Learning of efficient backprojections}
\label{sec:learn_of_bw_weights}

We describe our theory in a rate-based coding scheme.
\deleted{Following the convention defined for artificial neural networks, different cortical areas are represented by layers.}
\deleted{The somatic potentials of all neurons follow the dynamics of leaky integrators:
given an input current $\bm I[t]$, the membrane potential $\bm u$ obeys $C_m \bm {\dot{u}} = -\gl \, \bm u + \bm I[t]$, where $C_m$ denotes the capacitance, and $\gl$ the leak conductance of the cell membrane.
These dynamics imply a delayed response of the somatic potential with membrane time constant $\taueff \coloneqq C_m/\gl$.
Our theory describes neural dynamics where the current $\bm I[t]$ contains a local error signal.}
\replaced{The architecture is modeled with layers $\ell = 1\, \ldots \, N$ of leaky-integrator neurons as}{For concreteness, but without loss of generality, we consider the leaky-integrator model for a layered architecture with $\ell = 1\, \ldots \, N$,}
\begin{align}
	\label{eq:LI_base_model}
	\taueff \bm {\dot u}_\ell = - \bm u_\ell + \bm b_\ell + \bm W_{\ell,\ell-1} \bm r_{\ell-1} + \bm e_\ell + \bm \xi_\ell \,.
\end{align}
In this description, the somatic potential $\bm u_\ell$ integrates neuron bias $\bm b_\ell$, the bottom-up input rate $\bm r_{\ell-1}$ weighted with $\bm W_{\ell,\ell-1}$, and the local error $\bm e_\ell$.
\added{The integration time scale is set by the effective membrane time constant $\taueff$, defined by conductance and capacitance of the cell.}
\added{An essential ingredient is the} noise component $\bm \xi_\ell$.

\added{
	Our theory integrates the prospective coding mechanism of Latent Equilibrium~\cite{haider2021latent},
	which solves the relaxation problem of slow physical substrates, which disrupts inference as well as learning (see Methods).
	This is achieved by calculating the neural output from the prospective voltage $\bm {\breve u}  \coloneqq  \bm u + \taueff \,\frac{d\bm u}{dt} $ for all neurons.
	As a result, the rates $\bm r_\ell \coloneqq \varphi(\bm \ubreve_\ell)$ follow the neuron's inputs quasi-instantaneously.
}

The central question of cortical credit assignment is how the error $\bm e_\ell$ is calculated, given an error signal in a higher area,~$\bm e_{\ell+1}$, and how this error signal is used to adjust bottom-up weights.
Plenty of solutions to this question have been proposed~\cite{10.3389/fncom.2016.00094}.
Here, we focus on theories which can be formulated such that forward weights are updated as $\bm {\dot{W}}_{\ell,\ell-1} \propto  \bm e_\ell \, \bm r_{\ell-1}^T$.
Under this scheme, several theories for bio-plausible error transport exist~\cite{10.1162/089976603762552988,scellier2017equilibrium,haider2021latent,lee2015difference,sacramento2018dendritic,meulemans2020theoretical}\deleted{, among them contrastive Hebbian learning or difference target propagation}.
They have in common that errors in a higher layer $\bm e_{\ell+1}$ are propagated down through feedback (top-down) weights $\bm B_{\ell,\ell+1} $ to form errors in a given layer~$\bm e_{\ell}$.

Typically, \replaced{symmetric weights are}{a symmetric overall weight matrix is} assumed, such that $\bm B_{\ell,\ell+1} = [\bm W_{\ell+1,\ell}]^T$~\cite{10.1162/089976603762552988,scellier2017equilibrium,sacramento2018dendritic,NEURIPS2020_1abb1e1e}.
\added{Alternatively, learning schemes apply $\bm \dot{\bm{B}}_{\ell,\ell+1} =[ \bm \dot{\bm{W}}_{\ell+1,\ell}]^T$  \cite{374486,payeur2021burst,roelfsema_attention-gated_2005,pozzi2018biologically}, leading to weight alignment.}
\replaced{Both assumptions relate}{This assumption relates} the above \added{error propagation} schemes to classical \deleted{error }backpropagation\deleted{ based on gradient descent on a loss function}, where $\bm e_\ell = \varphi' \cdot [\bm W_{\ell+1,\ell}]^T \bm e_{\ell+1}$%
.
However, this assignment of weights \added{(or weight updates)} implies that top-down synapses in layer $\ell$ must adapt to the \deleted{(potentially distant) }bottom-up synapses in layer $\ell+1$.
This issue of how two spatially distant synapses (e.g.~across cortical areas) can \replaced{maintain similar weights}{keep up a similar weight} when one of them is learning is known as the \textit{weight transport problem}.

A proposed solution is \replaced{replacing}{the replacement of} $ \bm B_{\ell,\ell+1}$ with a random, fixed weight matrix (known as feedback alignment, FA~\cite{lillicrap2014random}).
However, FA has been shown to solve credit assignment inefficiently, and does not scale well to complex problems~\cite{nokland2016direct,moskovitz2018feedback,bartunov2018assessing,lansdell2019learning}.
We are therefore motivated to learn top-down weights such that credit assignment is improved compared to random feedback with layer-wise connections.
Furthermore, we would like to learn all weights $\bm B_{\ell,\ell+1}$ simultaneously, and in a way which does not interrupt feed-forward inference or learning of bottom-up weights.

\vspace{.25cm}

\begin{figure}[t]
	\centering
	\includegraphics[height=5cm]{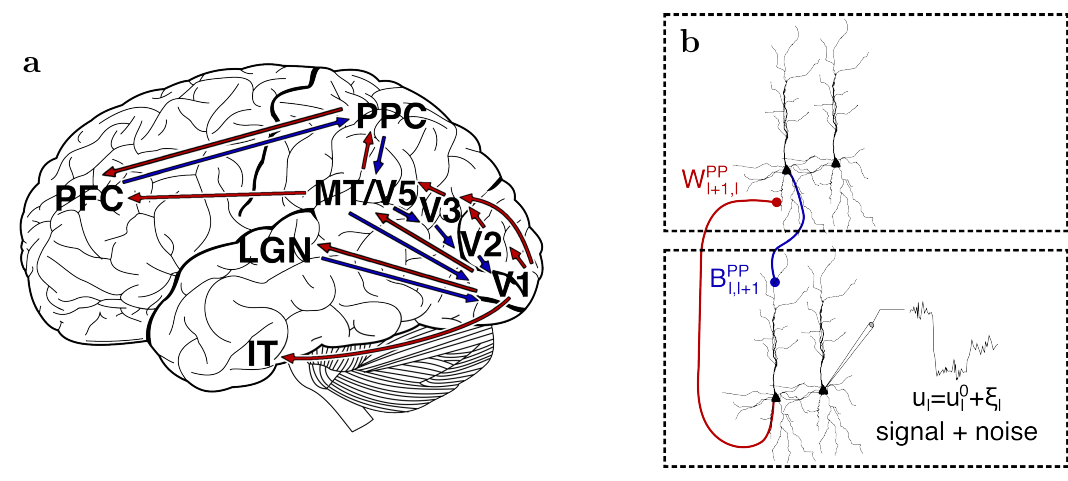}
	\caption{
		\textbf{Sensory processing over cortical hierarchies.}
		\textbf{a:}~Brain areas in the visual pathway beyond the primary visual cortex (V1).
		Information is propagated to higher areas (red arrows) such as V2, V4, the medial temporal (MT) area, and beyond.
		In order to assign credit, feedback information from higher level areas needs to be propagated top-down (blue arrows). Adapted from~\cite{archer2020temporal,visual_stream}.
		\textbf{b:}~Pyramidal cells as functional units of sensory processing and credit assignment.
		Top-down and bottom-up projections preferentially target different dendrites.
		Due to stochastic dynamics of individual neurons, noise is added to the signal.
	}
	\label{fig:general_setup}
\end{figure}

The general method we propose is explained in the following (see Fig.~\ref{fig:general_setup}).
An input signal $\bm r_0(t)$ (`data') is presented to the neurons in the lowest layer
for the duration of a presentation time $\Tpres$.
\deleted{This signal is propagated forward from layer $1$ to $N$, where each neuron follows the dynamics of Eq.~\eqref{eq:LI_base_model}.}
To reflect the inherent stochasticity of biological neurons subject to synaptic noise, thermal activity and probabilistic firing,
high-frequency noise $\bm \xi_\ell$ with a time constant $\tauxi$ is modeled at every neuron.
This noise is accumulated across layers, and propagated on top of the data signal. %
The top-down projections carry this mixed signal back to the lower layers, and we exploit the auto-correlation between noise signals to learn the corresponding feedback synapses\deleted{ based on a local alignment loss}.

\deleted{
	Concretely, for each hidden layer we sample Ornstein-Uhlenbeck noise $\bm \xi_\ell$ with zero mean and a small amplitude compared to the somatic potential $\bm u_\ell$ (see Methods, Eq.~\eqref{eq:SDE_noise}).
\added{
This type of noise is a natural choice to describe noisy dynamics of slow membranes, as it models neuron dynamics with Poisson input and leaky integration~\cite{ricciardi1979ornstein,petrovici2016stochastic}
}
\replaced{The}{This generic} noise term is added as a current to the soma of each hidden layer neuron, where it adds to the data signal to form a noisy firing rate.
Therefore, the sampled noise changes faster than the data signal (i.e.,~stimulus).
This condition, $\Tpres \gg \tauxi$, ensures that data and noise are separable in frequencies.
}

To learn the backwards weights, simple and only local computations need to be performed\deleted{ by the backprojections}.
At every top-down synapse ${\bm{B}}_{\ell, \ell+1}$, a high-pass filtered rate~$\widehat{\bm r}_{\ell+1}$ is computed, which extracts the noise signal; \replaced{learning rules using high-pass filtered signals can easily be implemented in neuromorphic hardware and have been observed in biology~\cite{bidoret2009presynaptic}}{this can be implemented in a physical substrate as the difference of the top-down rate with a low-pass filtered version}.
This filter separates the noise from the data portion of the top-down rate.
\deleted{
We learn the feedback synapses through minimization of a layer-wise alignment loss defined as
}
\deleted{
Minimization of $\mathcal{L}^\mathrm{PAL}_\ell$ leads to approximate alignment of $\bm B_{\ell,\ell+1}$ with $\bm W_{\ell+1,\ell}$, as detailed below.
Performing gradient descent on the alignment loss defines the top-down weight updates,}
\added{Feedback synapses are learnt through}
\begin{align}
	\bm \dot{\bm{B}}_{\ell,\ell+1} & =  \eta^\text{bw}_{\ell}  \big[\bm \xi_\ell \; \big( \widehat{\bm r}_{\ell+1} \big)^T - \alpha \, \bm B_{\ell,\ell+1}  \big] \;,
	\label{eq:IL_dBPP_LDRL}
\end{align}
\added{where $\alpha$ is a constant\deleted{ defining the size of the regularizer}.}
This rule can be applied \textit{simultaneously} to all layers to learn all backprojections $\bm{B}_{\ell,\ell+1}$, while allowing the learning of forward weights $\bm{W}_{\ell+1,\ell}$ at the same time.
Due to this crucial property, we name \replaced{our}{the above} method \textit{phaseless alignment learning} (PAL).
Note also that the learning \replaced{requires only}{rule is constructed solely from} information which is available pre- and post-synaptically for each neuron at each point in time\replaced{, }{.
This is} in line with our requirement of physical information processing, as well as phenomenological models of plasticity~\cite{clopath_connectivity_2010,bono_modeling_2017}.

\deleted{
Useful alignment of $\bm{B}_{\ell,\ell+1}$ \deleted{through minimization of $\mathcal{L}^\mathrm{PAL}_\ell$ }occurs in the following way (see also Methods).
At a given top-down synapse, the rate ${\bm r}_{\ell+1}$ arrives from the layer above.
Note that this rate is made from data as well as noise accumulated from all layers;
among this is also the noise originating in layer $\ell$.
The top-down synapse now calculates the high-pass filtered rate $\widehat{\bm r}_{\ell+1}$, discarding the data portion of the incoming signal.
As has been pointed out in Ref.~\cite{meulemans2021credit}, we then can exploit that the autocovariance of Ornstein-Uhlenbeck noise decays exponentially in time~\cite{sarkka_solin_2019}.
Therefore, the only non-zero correlation between all noise signals contained in~$\widehat{\bm r}_{\ell+1}$ and the current, local noise sample~$\bm \xi_\ell$ is proportional to the expectation value of the local noise auto-covariance $\bm \xi_\ell(t+\Delta t) \, \bm \xi_\ell(t)^T$.
Here $\Delta t$ denotes the time it takes for a noise sample to travel in a loop containing the layer above.
}

\replaced{For a given input sample and fixed bottom-up weights, the backwards weights align to}{Minimization of $\mathcal{L}^\mathrm{PAL}_\ell$  for a given input sample and fixed bottom-up weights aligns the backwards weights as}
\begin{align}
	\label{eq:BPP_converged_IL}
	\bm  B_{\ell,\ell+1}  \propto \varphi'(\bm \ubreven_\ell) \, \big[ \bm W_{\ell+1,\ell} \big]^T  \varphi'(\bm \ubreven_{\ell+1})\;,
\end{align}
where we refer to Methods for \replaced{a detailed explanation and}{the} derivation.
More generally, in a fully dynamical system with changing input, $\bm  B_{\ell,\ell+1}$ will converge to a weight which also aligns approximately with~$[\bm W_{\ell+1,\ell}]^T$, but is a mean over input data, i.e.~$\bm B_{\ell,\ell+1}  \propto \E[\varphi' \, [\bm W_{\ell+1,\ell}]^T \, \varphi']_{\bm r_0}$.

\deleted{
Note that our mechanism is able to take full advantage of the property of arbitrarily fast propagation due to Latent Equilibrium.
Noisy rates are calculated from the prospective voltage, and therefore the time delay between the top-down noise signal and the post-synaptic noise sample can become arbitrarily small.
This means that the correlation time scale of the Ornstein-Uhlenbeck noise $\tauxi$ can also be small, leading to fast convergence of backprojections \added{(the exact requirement on the time scales is $\tauxi \gtrsim \Delta t$)}.
In comparison, methods without prospective coding require~$\tauxi \gg \taueff$, such that top-down weights converge slowly (e.g.~\cite{meulemans2021credit}).
}

Learning backward weights in our framework is not disturbed by simultaneous learning of forward weights due to the frequency separation of data and noise:~as we require $\tauxi \ll \Tpres$, the error signal for forward weights can be recovered from backprojections by a low-pass filter with time constant larger than $\tauxi$. %
\deleted{Furthermore, PAL is also able to learn useful backprojections in absence of a teaching signal, facilitating efficient learning once an instructive signal is (re-)introduced.
In particular, top-down weights do not decay to zero if forward weights are kept fixed, even though the weight decay term $\bm {\bm \dot{B}}_{\ell,\ell+1} \propto - \alpha \, \bm B_{\ell,\ell+1}$ might suggest so.
The reason for this is that the expectation value of top-down weights, Eq.~\eqref{eq:BPP_converged_IL}, are formed from a balance between noise and the contribution due to the regularizer; see simulation results, Fig.~\ref{fig:teacher_student}, and Methods for details.
}

\subsection{Cortical microcircuit implementation}

\label{subsec:cort_mc}

\begin{figure}[t]
	\centering
	\includegraphics[width=.8\textwidth]{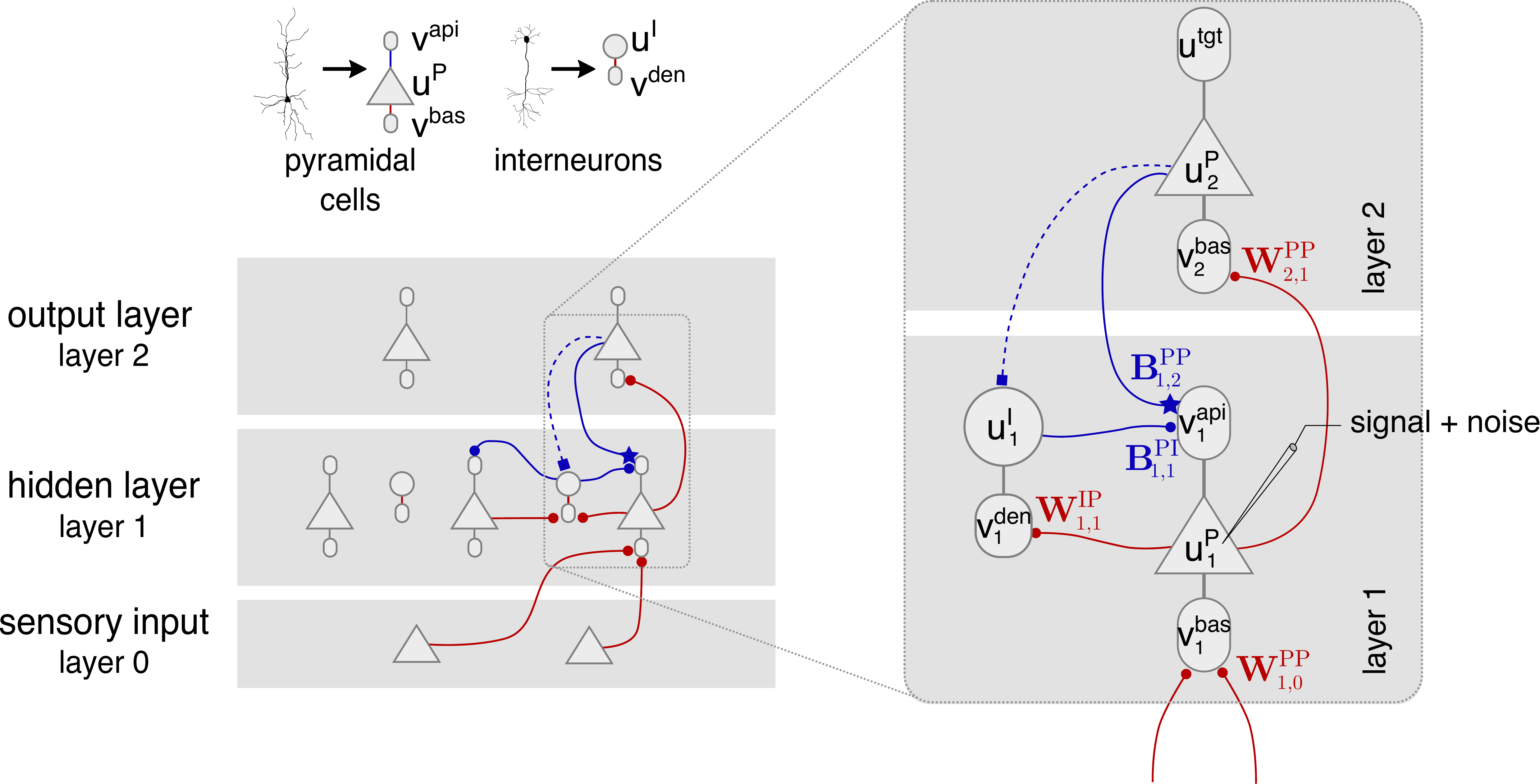}
	\caption{
		\textbf{Cortical microcircuit setup with one hidden layer.}
	\textbf{Left:} Full network with pyramidal cells and interneurons.
	Triangles represent somata of pyramidal neurons, with attached basal and apical compartments.
	Interneuron somata (circle) receive input from a single dendritic compartment and a nudging signal from a matching pyramidal cell in the layer above.
	\textbf{Right:} Single microcircuit.
	Somatic voltages contain bottom-up data signal, top-down error, and noise.
	The top-down synapses adapted with PAL are marked with a star.
}
	\label{fig:mc_setup}
\end{figure}

\added{PAL is compatible with a wide range of neuronal models: the only requirement is that they implement a variant of the dynamics of Eq.~\ref{eq:LI_base_model}.}
\added{For concreteness, }we now consider a particular implementation of PAL in the framework of dendritic cortical microcircuits~\cite{sacramento2018dendritic}.
This model has been introduced with biological plausible (error) signal transport in mind.
Each microcircuit is defined by \added{layer-wise connected} populations of two types of neurons, pyramidal cells and interneurons\added{, see Fig.~\ref{fig:mc_setup} and Methods}.
\deleted{These are organized in layers corresponding to cortical areas with a biologically plausible connection scheme, see Fig.~\ref{fig:mc_setup}} \added{and Methods}.

\deleted{
In absence of a teaching signal, pyramidal cells take the role of representation units, reflecting feed-forward activation.
They receive bottom-up information onto their basal dendrites and top-down activity in the distal apical dendrite, integrating both signals in the soma in accordance with observations of layer 2/3 pyramidal cells~\cite{jordan2020opposing}.
Pyramidal cells are modeled with a simplified three-compartment model with distinct basal, apical and somatic voltages~\cite{kording_supervised_2001,spruston_pyramidal_2008}.

The interneurons in this model are present in the hidden layers, and aim to represent a copy of the activation of pyramidal cells in the layer above.
Across layers, populations of pyramidal neurons and interneurons are arranged such that the number of interneurons in the hidden layers matches that of pyramidal cells in the layer above.
Interneurons are modeled with two compartments, representing dendritic tree and soma.
They receive lateral input from pyramidal cells in the same layer, and project back laterally to the same neurons.

The dynamics of the somatic membrane potentials of pyramidal cells $\bm u^\text{P}_\ell$ with $\ell=1, \,\ldots\, ,N-1$ are an instance of the general leaky-integrator equation~\eqref{eq:LI_base_model},
}
\deleted{
where Ornstein-Uhlenbeck noise $\bm \xi_\ell$ is modeled at all hidden layers, and $\bm \El$ denotes the leak potential.
A target signal can be introduced by clamping the apical compartment of the top layer pyramidal neurons to the target voltage $\bm u^\text{tgt}$.
Somatic voltages are determined by leaky integration of input basal and apical currents.
Dendritic compartment voltages are calculated instantaneously from their rate input, through~$\bm v^\bas_\ell = \bm \WPP_{\ell, \ell-1} \bm \rP_{\ell-1}$ for basal (bottom-up) input, $\bm v^\den_1 = \bm \WIP_{1, 1} \, \bm \rP_1 $ for the lateral input from pyramidal cells to interneurons, and the apical compartment potential determined from the sum of top-down and lateral activity, $\bm v^\api_1 = \bm \BPP_{1, 2} \bm \rP_{2} + \bm \BPI_{1, 1} \, \bm \rI_1$.
We refer to Methods for details.

As shown in Ref.~\cite{sacramento2018dendritic}, the weight updates in this model approximate those of error backpropagation in the limit of weak nudging (small top-down conductances):
}
\deleted{
where $\bm \vbashat_{\ell} \coloneqq \frac{\gbas}{\gl + \gbas + \gapi} \bm  v^\text{bas}_{\ell}$ denotes the conductance-weighted feed-forward input to each pyramidal cell, $\bm e_N \coloneqq \bm u^\text{tgt} -\bm \vbashat_{N} $ the output layer error, and a small parameter $\lambda$, which regulates the amount of top-down nudging.
}

In contrast to the model defined by Ref.~\cite{sacramento2018dendritic}, which employs \replaced{FA}{fixed feedback connections \added{$\bm \BPP_{\ell, \ell+1}$}}, we learn the backward connections using PAL with the scheme defined in Sec.~\ref{sec:learn_of_bw_weights}.
We consider noise in the hidden layers with a small amplitude compared to the corresponding somatic potentials.
Synaptic plasticity of thus-far fixed weights $\bm \BPP_{\ell,\ell+1}$ is enabled through the learning rule Eq.~\eqref{eq:IL_dBPP_LDRL}.
Finally, in order to preserve learning of feed-forward weights, we endow the update rule of $\bm \WPP_{\ell, \ell-1}$ with a low-pass filter with time constant $\taulo$.
Note that all computations required for PAL can be performed locally by the corresponding synapse.

\deleted{Additionally, we implement Latent Equilibrium into the microcircuit model as in Ref.~\cite{haider2021latent} by replacing all rates calculated from somatic potentials with rates obtained from the prospective voltage, $\bm \rPI_{\ell} = \varphi(\bm { u}^\text{P/I}_{\ell}) \mapsto \varphi(\bm {\breve u}^\text{P/I}_{\ell})$.
	This affects all compartment potentials as well as synaptic plasticity rules.}

\subsection{Experiments}

\deleted{We perform several experiments in order to evaluate PAL.
The base algorithm given by Eq.~\eqref{eq:IL_dBPP_LDRL} is applicable to rate-based neuron models.
Here, we focus our experiments on a microcircuit implementation as defined in the previous section, in order to demonstrate its merits as a bio-plausible method for learning.
We show that PAL is able to align top-down weights to useful backprojections, and compare weight updates to those of an ANN trained with BP.
A simple toy task (teacher-student) illustrates where PAL improves on \replaced{FA}{using fixed random backprojections}.
Using computer vision benchmark tests, we demonstrate that PAL is able to scale to bigger networks and more complex tasks.
Finally, we show that PAL facilitates credit assignment in deep networks, where multiple hidden layers are required for successful learning.
}

\deleted{
We stress that all simulations are performed with fully recurrent dynamics described by Eqs.~\eqref{eq:LI_base_model} and~\eqref{eq:mc_dyn},
differentiating our work from similar studies where the dynamics are replaced by steady-state approximations and the recurrency is implicitly removed by calculating separate forward and backward passes~\cite{sacramento2018dendritic,greedy2022single}.
All simulation parameters are given in Supplementary Information.
}

\subsubsection{Phaseless backwards weight alignment}

We first demonstrate that PAL aligns top-down weights in cortical microcircuits with the theoretical result given by Eq.~\eqref{eq:BPP_converged_IL}.
We simulate the dendritic microcircuit model with three hidden layers, keeping the forward weights~$\bm \WPP_{\ell, \ell-1}$ fixed while modeling noise in all hidden layers and learning all~$\bm \BPP_{\ell, \ell+1}$ simultaneously.
\deleted{In this experiment, we \added{(initially)} present no target to the output layer.}
Top-down weights as well as lateral weights from interneurons to pyramidal cells are adapted fully dynamically during training.

\begin{figure}[t]
	\centering
	\includegraphics[width=1.0\textwidth]{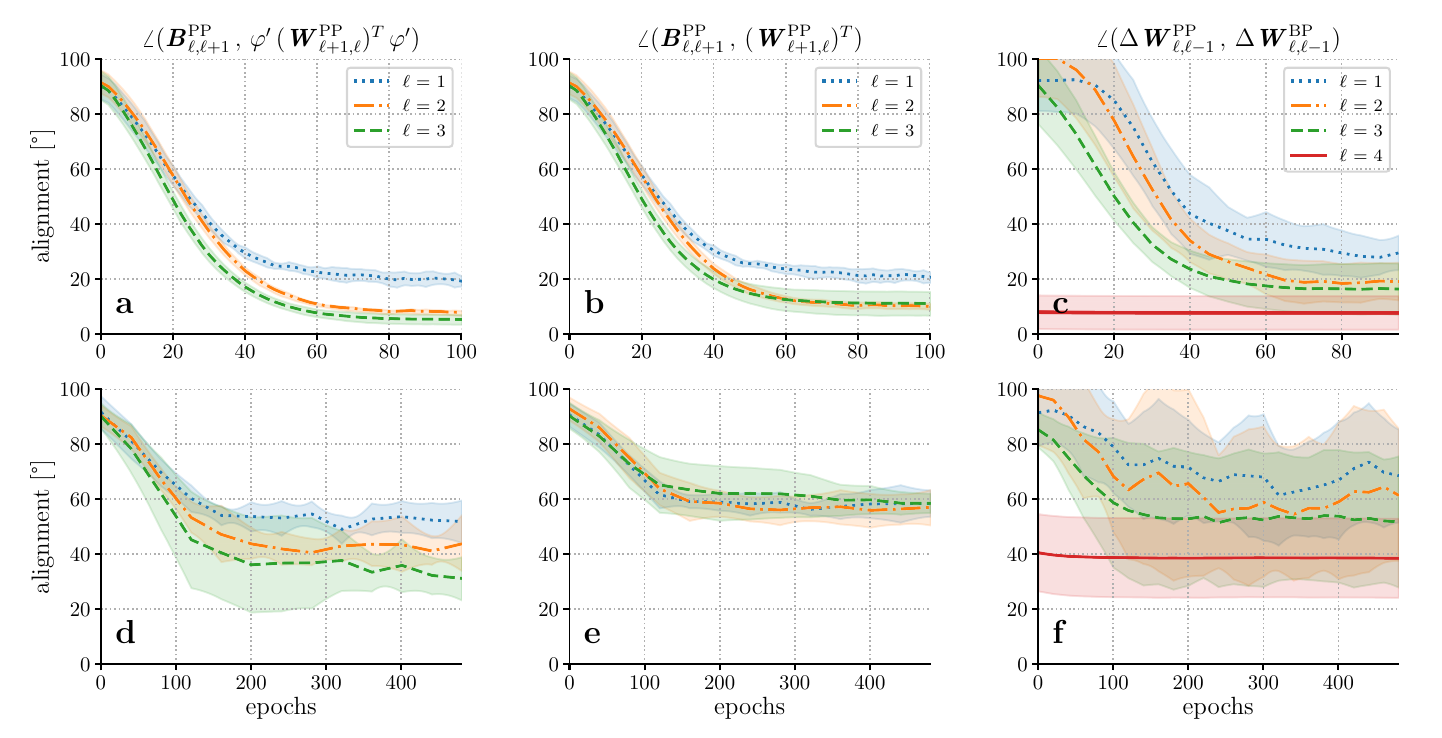}
\caption{
	\textbf{PAL aligns weight updates with backpropagation in deep networks.}
    \textbf{a, b, c:} We train the backward projections in a deep microcircuit network with layer sizes~[5-20-10-20-5] and sigmoid activation with no target present.
	All backward weights $\bm \BPP_{\ell,\ell+1}$ are learned simultaneously, while forward weights are fixed.
	Lines and shading show mean and standard deviation over \replaced{10}{ 5} seeds.
	Weights are initialized as $\bm \WPP \sim \mathcal{U}[-1,1]$, such that neurons are activated in their  linear regime.
	The right column compares the \textit{potential} forward weight updates generated from backpropagation using $\bm \BPP_{\ell, \ell+1}$ in the microcircuit model to those in an ANN with BP (see main text and Methods)\added{, where the instructive signal is provided by a teacher network with arbitrary forward weight configuration}.
	\textbf{d, e, f:} Same as above, but with weights initialized in non-linear regime, $\bm \WPP \sim \mathcal{U}[-5,5]$.
	Weight updates (f) are biased towards misalignment due to the dendritic microcircuit model, see Methods.
}
\label{fig:bw_learning}
\end{figure}

Results are shown in Fig.~\ref{fig:bw_learning}, where the upper and lower row correspond to the two cases where neurons are active in their linear/non-linear regime.
It is of interest to evaluate both of these regimes, as complex tasks cannot be solved with a fully linear network. Nevertheless, it is also not the case that all neurons are in the non-linear regime for all inputs; typically, there is a mixture of both states present in the network.

The first column shows the angle between top-down weights and Eq.~\eqref{eq:BPP_converged_IL}, demonstrating good agreement with the theoretical expectation over all hidden layers.
These backward weight configurations are useful, as they approximately align with the transpose of the forward weights; we show the corresponding alignment angle in the second column.
Alignment of top-down weights with the transpose is much better in the linear regime, as $\varphi' = \id$.
\added{We hypothesize that the larger misalignment angle in the non-linear case is due to the data-specific learning of backwards weights (see Discussion).}

In the third column, we show that errors propagated in a microcircuit with PAL approximately align with backpropagation.
After each epoch of training the backprojections, we evaluate the model in its current state by introducing a teaching signal.
This generates an error which propagates to all layers, and from which a forward weight update $\Delta \bm \WPP_{\ell, \ell-1}$ is constructed; see Methods for details.
\deleted{We stress that here, no weight update is applied, as in this experiment, we only demonstrate learning of top-down weights.}
We compare $\Delta \bm \WPP_{\ell, \ell-1}$ of this microcircuit model to the weight updates $\Delta \bm W^\text{BP}_{\ell, \ell-1}$ in an ANN with backpropagation and \replaced{equal}{equivalent} feed-forward weights.
The results \deleted{in the third column} demonstrate that PAL is able to propagate useful error signals through alignment of backward weights.

\subsubsection{Teacher-student setup}
\label{sec:ts}

To further demonstrate that PAL enables propagation of useful error signals, we turn to a simple teacher-student task.
Plasticity is \added{now} enabled in \replaced{forward}{all} synapses\deleted{, i.e.~forward weights are adapted}, too.
A microcircuit model consisting of a chain of two neurons is trained with PAL and\replaced{ FA, and the ideal case of BP}{, for comparison, random fixed backwards weights}.
\deleted{For comparison, we also show the ideal however bio-implausible case corresponding to BP, where top-down weights are set to the same value as the forward weights to the output neuron.}
A teaching signal is obtained from a similar two-neuron chain connected with fixed\replaced{,}{ and} positive weights.
\deleted{The teacher chain produces a non-linear input-output mapping determined by the choice of its synaptic weights.}
The task of the student is to adapt its weights to reproduce \replaced{the}{this} input-output relationship.

\begin{figure}[t]
	\centering
	\includegraphics[width=1.0\textwidth]{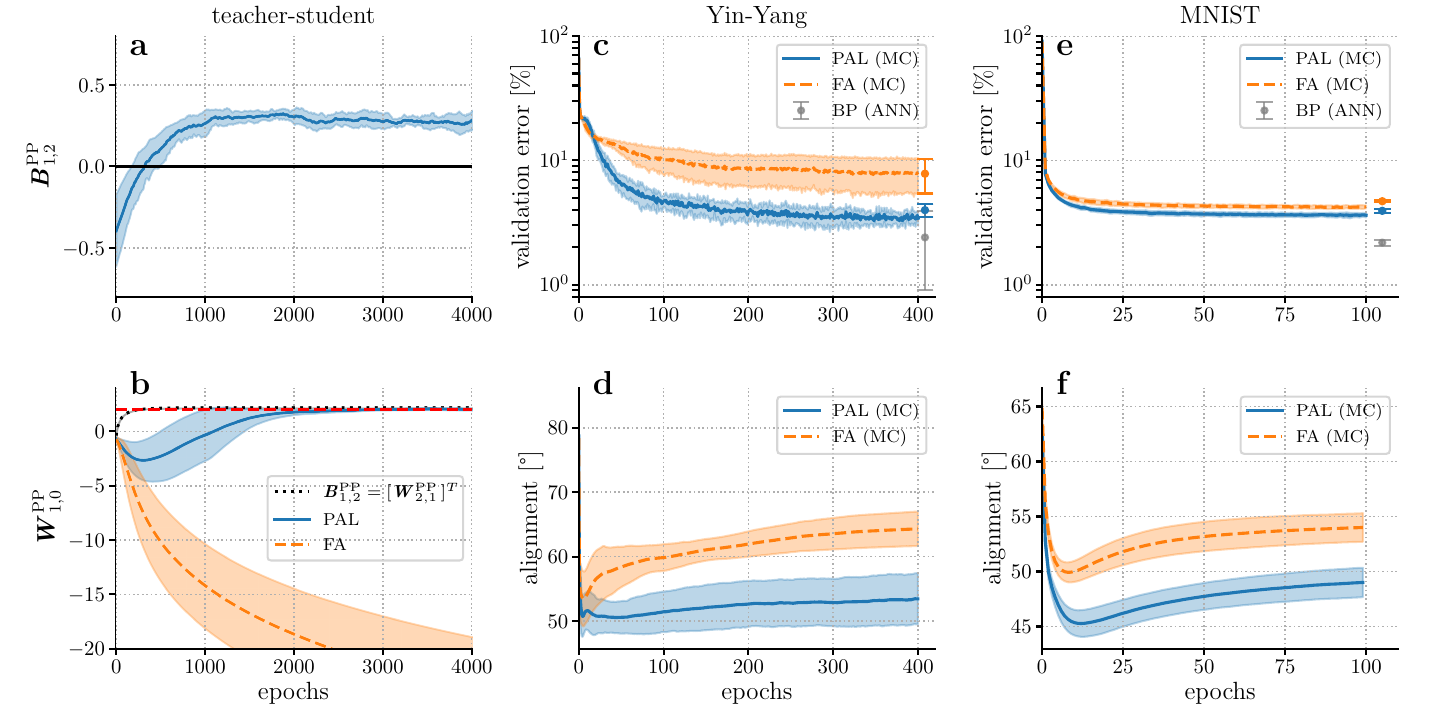}
\caption{
\textbf{PAL improves learning on teacher-student and classification tasks compared to fixed random synaptic feedback.}
\textbf{a, b:}
A chain of two neurons learns to mimic a teaching signal.
The student neurons are initialized with negative weights and need to flip the sign.
In particular, in order to achieve correct weights to the hidden neuron $\bm \WPP_{1, 0}$, positive feedback weights are required.
The teacher (red) has a positive weight.
    Results for student neuron chains are shown for PAL (blue) and FA (orange). The shading indicates mean and standard deviation over \replaced{10}{5} seeds.
Due to the wrong sign of the transported error, random synaptic feedback fails to solve the task, and weights diverge.
The models trained with PAL start in a similar fashion; however, after sign flip (at about 500 epochs, see \textbf{a}), the error signal becomes useful, and $\bm \WPP_{1, 0}$ converges to the weight of the teacher.
As a control, we also show the ideal solution with weight transport, $\bm \BPP_{1,2} =  (\bm \WPP_{2,1})^T$.\\
\textbf{c, d:}
Validation error during training and test errors for the Yin-Yang task (\textbf{c}) of the microcircuit model (MC) with network size [4-30-3].
For reference, we also show the test error in an ANN trained with BP with equal network size.
The shading indicates mean and standard deviation over 10 seeds.
\textbf{d:} Alignment angle between backwards weights $\bm \BPP_{1,2}$ and $ (\bm \WPP_{2,1})^T$.
While FA relies on alignment of forward weights, PAL improves on this by aligning the backward weights with the transpose of forward weights.\\
\textbf{e, f:}
Same as center column, but for the MNIST data set with network size [784-100-10].
\added{Note that the increased performance cannot simply be attributed to the inclusion of noise, see Supplement.}
}
\label{fig:teacher_student}
\end{figure}

In order to highlight an important shortcoming of \replaced{FA}{fixed feedback weights}, we initialize the student models with negative forward and backward weights.
As shown in Fig.~\ref{fig:teacher_student}, a model trained with \replaced{FA}{fixed random feedback weights} is not able to reproduce the teacher output and even has diverging weights.
This is caused by the wrong sign of the top-down weights: a positive error on the output layer is projected backwards through the negative synapse $\bm \BPP_{1, 2}$.
Thus, the weight $\bm \WPP_{1, 0}$ to the hidden layer grows negatively, further increasing the disparity between teacher and student weights.
PAL resolves this issue by approximately aligning $\bm \BPP_{1, 2}$ with $\bm \WPP_{2,1}$, thereby learning backwards weights with correct sign (left column in Fig.~\ref{fig:teacher_student}).
Note that as as long as $\bm \BPP_{1, 2}$ has not yet aligned, $\bm \WPP_{1,0}$ moves in the wrong direction, but as soon as $\bm \BPP_{1, 2}$ switches sign \deleted{(at epoch $\sim 500$)}, the forward weight is able to learn correctly.
\deleted{
\added{This also requires the weights $\bm \WPP_{2,1}$ to be positive, which is achieved independently of the method (PAL/FA) through learning of the top-layer (not shown).}
}

\subsubsection{Classification experiments}
\label{sec:classification}

We now turn to more complex tasks and evaluate PAL on classification benchmarks, while still working with the biologically plausible microcircuits as the base model.
Due to the complexity of simulating microcircuit models with full dynamics, we focus this evaluation on the computationally effective Yin-Yang task, and perform experiments on MNIST digit classification as a sanity check.

The Yin-Yang classification problem~\cite{kriener2021yin} is designed to be a computationally inexpensive task which nonetheless requires useful error signals to reach the lower layer of a network, i.e.~is able to differentiate the error propagation quality between FA and BP or variants of it.
\deleted{The task consists in learning to map 2d input coordinates correctly to three distinct categories.}
ANNs trained with backprop can solve this task with as few as 30 hidden neurons (test error $2.4 \pm 1.5$~\%)~\cite{kriener2021yin}.
\deleted{This requires the formation of a useful hidden layer representation, which is more likely if backwards weights are adapted instead of random and fixed.}

The microcircuit models with PAL achieve a test error of $4.0 \pm 0.4$~\%, performing considerably better than microcircuits with fixed random feedback weights at $7.8 \pm 2.4$~\% (Fig.~\ref{fig:teacher_student}, center column).
This is reflected in the increased alignment between the transpose of forward and backward weights.

We also perform the MNIST \deleted{digit classification }task with a similar setup (right panel in Fig.~\ref{fig:teacher_student}).
In a similar vein to the Yin-Yang experiments, this single and small hidden layer is chosen as to highlight whether a good latent representation is formed.
We achieve a final test error $3.9 \pm 0.2$~\% using PAL and $4.7 \pm 0.1$~\% \deleted{with microcircuits} with FA.

We highlight that our results were obtained by simulating a fully dynamical, recurrent and bio-plausible system with weight and voltage updates applied at every time step.
This is in contrast to previous simulations using bio-plausible networks with recurrency, where simplified network dynamics were assumed for computational feasibility~\cite{sacramento2018dendritic,payeur2021burst,greedy2022single}.
\added{As a direct comparison, simulations on the MNIST task in Ref.~\cite{sacramento2018dendritic} use the steady state approximation of voltage dynamics, and weight updates are calculated in two distinct steps. In effect, this simplifies the recurrent dynamics defined by Eq.~\eqref{eq:mc_dyn} to those of an ANN with separate forward and backward phases and voltage buffering.}
Such approximations do not accurately reproduce the dynamics of recurrent physical networks, \added{and we highlight that our results represent a significant step towards efficient bio-realistic microcircuit simulations.}

\subsubsection{Efficient credit assignment in deep networks}
\label{sec:deep_nets}

The previous analyses have shown that PAL can learn useful backprojections in dynamical systems.
The simulations performed with microcircuit models stress the bio-plausibility of PAL. However, the microcircuit model (both with and without PAL) carries the issue that error signals decay with increasing hidden layer number (see Methods).
PAL is designed to learn useful backprojections in deep hierarchies; in order to demonstrate the full capability of our method, we now relax our requirement for bio-plausible error transport and shift away from the dendritic microcircuit model.

We revisit the general leaky integrator model defined by Eq.~\eqref{eq:LI_base_model}\replaced{, where}{.
The difference between this simpler model and the microcircuit model is the exclusion of interneurons.
Instead,} errors are transported directly via $\bm e_\ell = \varphi'(\bm \ubreve_{\ell}) \cdot \bm B_{\ell,\ell+1} \, \bm e_{\ell+1} $.

\begin{figure}[h]
	\centering
	\includegraphics[width=0.9\textwidth]{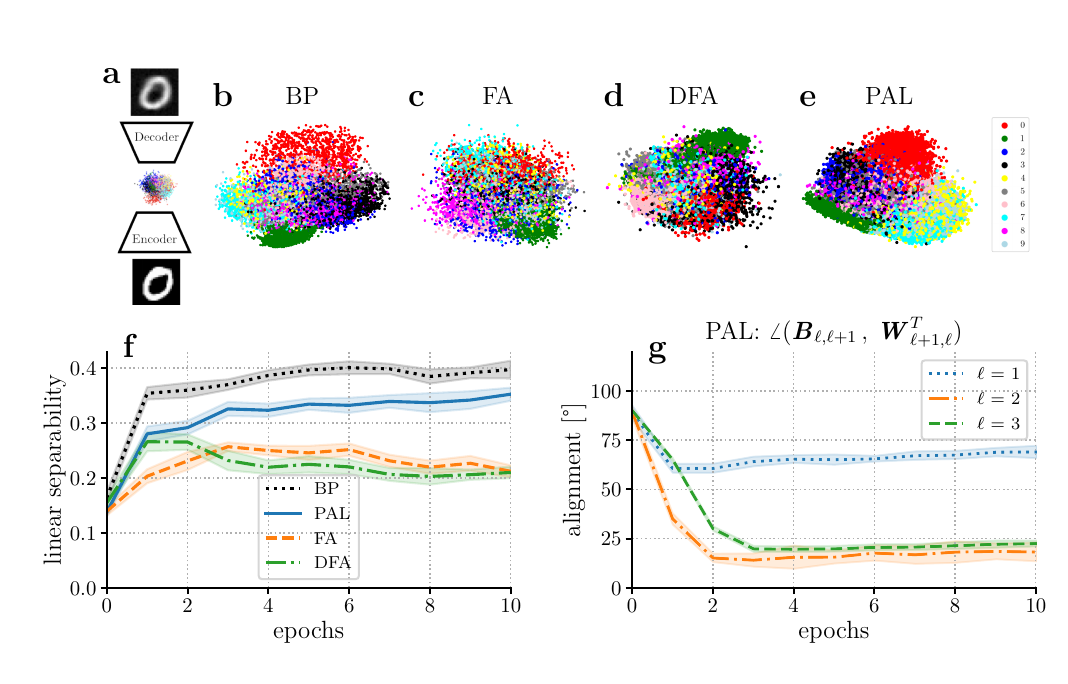}
	\caption{
		\textbf{PAL learns useful latent representations, where feedback alignment fails to do so.
		}\\
		\textbf{a:} Encoder/decoder setup network with size [784-200-2-200-784].
		MNIST digit dataset is fed into an encoder network with two output neurons.
		A stacked decoder network aims to reproduce the original input.
		\textbf{b-d:} Latent space activations after training.
		We show the activations after training in the two-neuron layer of one seed for all samples in the test set; colors encode the corresponding label.
		Backpropagation and PAL show improved feature separation compared to \added{(direct)} feedback alignment.
		\textbf{e:} Linear separability of latent activation.
		\textbf{f:} Alignment angle of top-down weights to all layers for the PAL setup (mean and standard deviation over \replaced{10}{5} seeds).
		PAL is able adapt top-down weights while forward weights are also learned.
		Note that the layer $\ell = 1$ maps 200 neurons onto two in the forward direction, leading to many possible solutions in the backwards direction, and hence a larger alignment angle is to be expected.
	}
	\label{fig:eff_deep_nets}
\end{figure}

We demonstrate the capability of PAL for credit assignment using the MNIST-autoencoder task (Figure~\ref{fig:eff_deep_nets})~\cite{lansdell2019learning}.
Autoencoders can be used to to greatly compress an input image to a latent representation, in this case reducing down to a vector of dimension two.
In order to decode such a representation, a successfully trained autoencoder network should show a separation of input of different classes in the latent space.
To learn a well separated latent representation, suitable error signals need to travel through the whole network to train the encoder weights.
\added{We compare the performance of PAL with FA and its variant direct feedback alignment (DFA), where the feedback signal from the output layer is sent directly to all hidden layers~\cite{nokland2016direct,crafton2019direct}.
}

\begin{figure}[h]
\centering
\includegraphics[width=.7\textwidth]{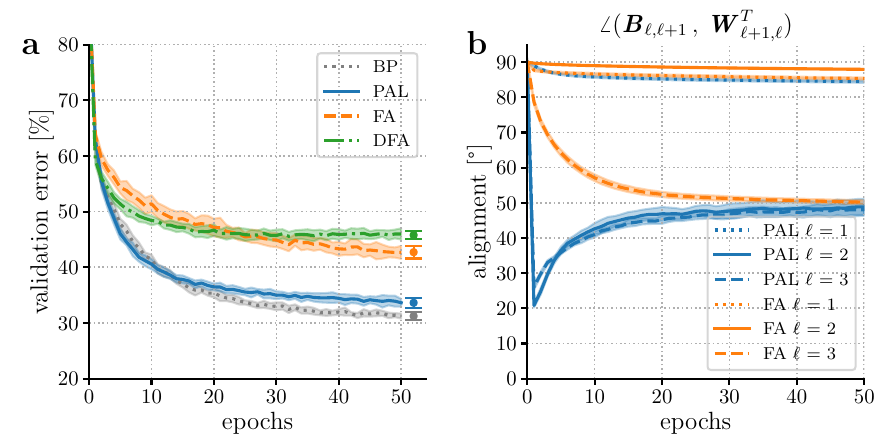}
\caption{
\textbf{PAL outperforms (D)FA on CIFAR-10 using a continuous-time CNN.}
	The network consists of four layers (Conv2d + MaxPool(2), Conv2d + MaxPool(2), dense, linear) with sigmoid activations, see Methods.
	\textbf{a:}~Validation and test accuracies show that PAL outperforms FA and DFA on CIFAR-10.
	Noise is disabled for PAL during evaluation (val+test) in order to enable a fair comparison.
	Test accuracies are:
	BP~$31.2 \pm 0.7~\%$;
	PAL~$33.9 \pm 0.7~\%$;
	FA~$42.7 \pm 1.1~\%$;
	DFA~$46.8 \pm 0.7~\%$.
	\textbf{b:}~Alignment between forward and backwards weights for PAL and FA for all layers.
	PAL is enabled for layers $\ell=2,3$ ($\ell = 1$ is a convolutional layer, for which PAL is not defined).
	All curves show mean and error over 10~seeds.
}
\label{fig:cifar_10}
\end{figure}

We evaluate the latent space separability by training a linear classifier.
Results shown in Fig.~\ref{fig:eff_deep_nets}~(e) demonstrate that networks trained with PAL achieve linear separability close to BP{ ($(35.2 \pm 1.2)$~\% vs.~$(39.7 \pm 1.6)$~\%, respectively)}, while training with \added{(direct)} FA leads to significantly poorer \replaced{feature separation}{linear separability}{ at a test accuracy of $(21.2 \pm 1.0)$~\% (DFA:~$(21.0 \pm 1.0)$~\%)}.
Fig.~\ref{fig:eff_deep_nets}~(f) shows that PAL is able to learn the transpose of forward weights across hidden layers.
\deleted{Our results imply that with PAL, the network is able to transport useful error signals and learn suitable weights throughout all hidden layers, whereas FA leads to poor feature separation.}

\added{Similarly, we show that PAL outperform (D)FA on CIFAR-10 using a continuous-time ConvNet (Fig.~\ref{fig:cifar_10}).
PAL clearly outperforms random feedback by learning weights to a high degree of alignment.}

\section{Discussion}

We have introduced PAL, a general method of learning backprojections in hierarchical, dynamical networks.
Our theoretical results and simulations show that PAL provides online learning of forward and backward weights in a phaseless manner.
As a general method, it is applicable to models where time-continuous activity is propagated, and approximately aligns feedback weights with those of backpropagation.
PAL fulfills the requirements of learning and signal transport in physical systems: all necessary information is available locally in time and space.

In our evaluation, we have emphasized the biological plausibility of PAL as a model of sensory processing.
PAL could be implemented in biological components; in particular, it explicitly exploits noise found in physical systems and makes use of simple filtering techniques for disentangling signal and noise where needed.
We argue that a cortical realization of PAL (or a variant) would be evolutionarily more advantageous than fixed feedback weights, as it implements a significantly more efficient solution to the weight transport problem.
\added{PAL requires top-down weight learning to be faster than bottom-up learning, such that error signals are transported correctly before they are used.
In cortex, this could potentially be regulated by the neurotransmitter acetylcholine (ACh), involved in bottom-up and top-down attention modulation~\cite{sato1987functional,soma2012cholinergic,kang2014boosting}.
Learning of top-down before bottom-up weights could be modulated through release of ACh in top-down information integration sites of neurons.}

Our simulation results show that PAL is able to outperform FA in terms of credit assignment in biologically plausible, recurrently connected networks.
The requirements for PAL are quite general, and we stress that the dendritic microcircuit model with PAL is only one possible implementation, which however is notable for its bio-plausible error transport.
In principle, it can be argued that a test error comparable to PAL could have been achieved with fixed random backprojections by scaling up the hidden layer size (not shown).
Nevertheless, it is advantageous for a method to perform well while requiring fewer neurons/synapses, for biological plausibility but also energy efficiency.
It has been suggested that the inefficiency of FA can mitigated through \replaced{DFA}{direct random backprojections (direct feedback alignment, DFA), where the feedback signal from the output layer is sent directly to all hidden layers, passing only through a single random feedback matrix~\cite{nokland2016direct,crafton2019direct}}\added{, for which we however do not find evidence}.
\replaced{Additionally,}{However,} in the context of cortical hierarchies, this presupposes skip connections from one higher cortical area to all lower areas instead of layer-wise connections.
While such connections have been observed in the cortex, the visual stream is largely organized in hierarchical manner, with significantly weaker correlation between non-neighboring areas~\cite{HAAK201873}.

However, like any algorithm based on learning from data samples, weights trained with PAL are data-specific.
That is, top-down weights are aligned with an average over many input samples $\bm r_0$ as $\bm B_{\ell,\ell+1}  \propto \E[\varphi' \, [\bm W_{\ell+1,\ell}]^T \, \varphi']_{\bm r_0}$,
leading to imperfect alignment of transported errors with those of backpropagation~(see Fig.~\ref{fig:bw_learning}\added{; in practice, weight alignment is restricted to at least 20$^\circ$}).
Lower learning rates $\eta^\text{bw}$ lead to an inclusion of more samples into the expectation value;
by sampling over the whole training set, data-dependency can be minimized -- however, as forward weights need to evolve slower than backward weights, this leads to slow overall learning.
In contrast to this, \cite{ernoult2022towards} circumvents this issue by separate phases of forward and backward learning for each data sample, but no fully on-line solution is currently known.
As we have shown, the error transported by this data-specific weights can still be efficient in learning to solve complex tasks.

Note that PAL is able to make full use of prospective coding -- i.e.,~that all information propagation occurs through prospective rates $\varphi(\bm {\breve{u}}_\ell)$, which converge to their steady state quasi-instantaneously.
\replaced{This}{In particular, this} ensures that \replaced{weight updates}{the weight update rules} are constructed from useful learning signals at all times, not only after the neuronal dynamics have settled into a steady state.
As a consequence, we were able to simulate our dynamical system with fully continuous voltage dynamics and learning of all synapses enabled at all times.
This is an important difference compared to previously known bio-plausible mechanisms of learning in dynamical systems, which have mitigated the issue of slow relaxation through slow or phased learning~(see e.g.~\cite{6796552,bengio2015early,sacramento2018dendritic,guerguiev2017towards,scellier2017equilibrium,mesnard2019ghost,10.1162/089976603762552988,Song2022.05.17.492325}), and/or by re-initializing the somatic potentials to their bottom-up input state for every data sample~\cite{scellier2017equilibrium,sacramento2018dendritic,meulemans2021credit}.

As our theory is based on a rate-based abstractions of neural dynamics, there remain several open questions of bio-plausibility and realism.
Extensions to our theory may implement Dale's law of either inhibitory or excitatory activity, a constraint which could be realized by separate populations of neurons~\cite{Cornford2020.11.02.364968}.
Equally, the dendritic microcircuit model could model biology more closely by implementing spiking neuronal output, and arguing that credit is assigned through a probabilistic interpretation.
\added{Future work will investigate how PAL can be related to theories of spike-based symmetrization~\cite{burbank2015mirrored}.}

Our results clearly point towards future work on theories of on-line learning in the cortex and on neuro-inspired hardware.
Building on similar ideas in the literature~\added{\cite{6797856,Rusakov2020,McDonnell2011,Faisal2008,xie_learning_2004,fiete2007model}}, we hypothesize a general principle of using noise in physical systems for learning, instead of considering it an undesirable side effect when modeling substrates.
PAL can serve as a blueprint for the greater philosophy of viewing noise as a resource rather than a nuisance.

\section{Methods}

As in the main text, bold lowercase (uppercase) variables $\bm x$ ($\bm X$) denote vectors (matrices).
The partial derivative of the activation given by $\bm r_\ell = \varphi (\bm {\breve{u}})$ is denoted by
$\varphi'(\bm {\breve{u}})$, which is a diagonal matrix with $\mu$-th entry $\frac{\partial r_\mu}{\partial  {\breve{u}_\mu} }$.

\subsection{Prospective Coding}

As neurons in our theory are modelled by leaky integrators, the somatic voltage follows the low-pass filtered sum of input currents, and therefore exhibits a slow response to its input.
This effect multiplies with increasing layer number, resulting in the requirement to present an input for many membrane time constants to allow both input signals from the bottom and learning signals from the top to fully propagate through the whole network.
Additionally, slow neuron dynamics do not only slow down the flow of information, but also introduce incorrect error signals, as demonstrated in Ref.~\cite{haider2021latent}.

Several schemes have been proposed to solve this issue.
A common fix is scheduled plasticity, where synapses are only learned once the system has settled into the equilibrium state~\cite{6796552,bengio2015early,sacramento2018dendritic,guerguiev2017towards,scellier2017equilibrium,mesnard2019ghost,10.1162/089976603762552988,Song2022.05.17.492325}.
This leads to slow learning, and the need to explain the phased plasticity through a biologically plausible mechanism.
In Ref.~\cite{haider2021latent} and our model however, the issues caused by response lag are overcome by calculating the firing rate of each neuron based on the prospective future voltage:
all neural outputs and weight updates are calculated from the prospective voltage $\bm {\breve u}  \coloneqq  \bm u + \taueff \,\frac{d\bm u}{dt} $.
\added{The inclusion of the temporal derivative adds \textit{`what neurons guess that they will be doing in the future'}~\cite{haider2021latent} to their current state.
Note that the effect of prospectivity (where the output of a neuron is a function also of $\bm{\dot{u}}$) can already be observed in standard neuron models~\cite{plesser2000escape}, and has been seen experimentally in Layer~5 pyramidal cells~\cite{10.1093/cercor/bhm235}; for a comprehensive explanation of biological plausibility, see Sec.~3 of Ref.~\cite{haider2021latent}.}

The implementation of prospective coding with PAL is essential for fast transfer of information, ensuring quick convergence of weights.
\added{%
Noisy rates are calculated from the prospective voltage, and therefore the time delay between the top-down noise signal and the post-synaptic noise sample can become arbitrarily small.
This means that the correlation time scale of the Ornstein-Uhlenbeck noise $\tauxi$ can also be small, leading to fast convergence of backprojections \added{(the exact requirement on the time scales is $\tauxi \gtrsim \Delta t$)}.
In comparison, methods without prospective coding require~$\tauxi \gg \taueff$, such that top-down weights converge slowly (e.g.~Ref.~\cite{meulemans2021credit}), see Fig.~\ref{fig:PAL_needs_LE} in Supplementary Information.
}

\subsection{Alignment of feedback weights}

We show how the weight transport problem can be solved through alignment of top-down weights.
We keep our description general by discussing the basic leaky integrator model with noise,
\begin{align}
	\taueff \bm {\dot u}_\ell &= - \bm u_\ell + \bm b_\ell + \bm W_{\ell,\ell-1} \bm r_{\ell-1} + \bm e_\ell + \bm \xi_\ell
\end{align}
with $\bm e_\ell$ propagated downwards from the upper layer error through feedback connections $\bm B_{\ell,\ell+1}$.
As detailed in Sec.~\ref{sec:learn_of_bw_weights}, we do not discuss error propagation itself, but focus instead on the learning of feedback weights from the rates generated within each layer.
The procedure to learn the backwards weights relies on two recent theoretical advancements:~the local difference reconstruction loss defined in Ref.~\cite{ernoult2022towards} trains backwards weights to approximate backpropagation, whereas the proposal of Ref.~\cite{meulemans2021credit} to consider Ornstein-Uhlenbeck noise enables us to learn all backwards weights simultaneously.

\added{
	Concretely, for each hidden layer we sample Ornstein-Uhlenbeck noise $\bm \xi_\ell$ with zero mean and a small amplitude compared to the somatic potential $\bm u_\ell$.
	This type of noise is a natural choice to describe noisy synaptic currents, as it models synapse dynamics with Poisson input and finite (non-zero) synaptic decay times~\cite{petrovici2016stochastic}.
	\replaced{The}{This generic} noise term is added as a current to the soma of each hidden layer neuron, where it adds to the data signal to form a noisy firing rate.
	The noise changes faster than the data signals (i.e.,~stimulus and learning signal).
	This condition, $\Tpres \gg \tauxi$, ensures that data and noise are separable in frequencies.
}

\added{
The variance of this noise can be modulated by the interplay between background firing rates and synaptic weights, while conserving its mean~\cite{ricciardi1979ornstein}, and is often used in the diffusion approximation for stochastic neurons~\cite{gerstner2014neuronal}. We note that a further interesting component is added when considering conductance-based synaptic interactions. Here, the stochasticity of neuronal membranes can depend non-monotonically on presynaptic rates, and counterintuitively decrease when rates increase~\cite{petrovici2016form}. A similar effect can be derived in the context of Bayesian inference~\cite{jordan2021learning} and has been observed experimentally~\cite{crochet2011synaptic}.
}

To begin, we model Ornstein-Uhlenbeck noise $\bm \xi_k$ in each hidden layer.
The noise signal is generated from low-pass filtered white noise,
\begin{align}
	\bm {{\dot{\xi}}}_\ell(t) = -\frac{1}{\tauxi} \big[ {\bm  \xi}_\ell(t) - \bm \mu_\ell(t)  \big]\,,
	\label{eq:SDE_noise}
\end{align}
with $ \bm \mu_\ell(t) \sim \mathcal{N}(0, \sigma^2)$, and low-pass filtering constant $\tauxi$ smaller than the usual presentation time $T_\text{pres}$ of input signals $\bm r_0$. Therefore, $\bm \xi_k$ represents an Ornstein-Uhlenbeck process with %
high frequency compared to the inference and error signals.
We choose the scale of noise $\sigma$ such that it is small compared to the somatic potential in all hidden layers.
If we denote as $\varphi(\bm \ubreven_k)$ the neuron output in absence of noise, we can expand to first order in small noise, $\bm r_k \approx \varphi(\bm \ubreven_k) + \varphi'(\bm \ubreven_k) \, \bm \xi_k$.
This signal is sent to the corresponding higher layer $k+1$ through the weight $\bm W_{k+1,k}$, where the upper layer noise $\bm \xi_{k+1}$ is added on top.
This continues through all layers up to the output layer.
Hence, for a given layer $\ell+1$, the output rate is modified by noise to be
\begin{align}
	\label{eq:app_rP_noise_expanded}
	\bm r_{\ell+1} \approx \varphi(\bm \ubreven_{\ell+1})  + \varphi'(\bm \ubreven_{\ell+1}) \,   \bm \xi_{\ell+1} + \varphi'(\bm \ubreven_{\ell+1}) \, \sum_{m=1}^{\ell} \big[ \prod_{n=m}^{\ell} \bm W_{n+1,n} \varphi'(\bm \ubreven_n) \big] \,   \bm \xi_m \;.
\end{align}
\added{Note that we omit noise originating in downstream areas ($\bm \xi_{\ell+2}, \ldots , \bm \xi_{N}$) as it quickly averages to zero (see below).}
The noise-inclusive rate $\bm r_{\ell+1}$ is also propagated top-down.
In the case of layer-wise feedback connections, $\bm r_{\ell+1}$ is sent to layer $\ell$ through the synapse $\bm B_{\ell,\ell+1}$.
Therefore, the information locally available to learn useful top-down weights is restricted to the pre-synaptic rate $\bm r_{\ell+1}$ and the post-synaptic potential including noise.

We aim now to learn the top-down synapses by exploiting the auto-correlation of noise.
The slow portion $\varphi(\bm \ubreven_{\ell+1})$ of the pre-synaptic signal is not useful for learning of backwards weights, as it contains correlations across layers which cannot be canceled.
Therefore, we extract the noise-induced portion of the signal with a high-pass filtered version of the top-down rate, $\widehat{\bm r}_{\ell+1}$~\cite{meulemans2021credit}.
The dynamics of $\bm B_{\ell,\ell+1}$ are derived from gradient descent on the local alignment loss $\mathcal{L}^\mathrm{PAL}_\ell$\added{, defined as}
\begin{align}
    \mathcal{L}^\mathrm{PAL}_\ell &= - \bm \xi_\ell^T (t) [\bm B_{\ell,\ell+1}  \widehat{\bm r}_{\ell+1}(t)] + \frac{\alpha}{2} \, \|\bm B_{\ell,\ell+1} \|^2\;.
    \label{eq:L_PAL}
\end{align}
Gradient descent yields the learning rule
\begin{align}
	\bm \dot{\bm{B}}_{\ell,\ell+1} \coloneqq - \eta^\text{bw}_{\ell} \, \nabla_{\bm{B}_{\ell,\ell+1}}  \mathcal{L}^\mathrm{PAL}_\ell  &= \eta^\text{bw}_{\ell}  \big[\bm \xi_\ell \; \big( \widehat{\bm r}_{\ell+1} \big)^T - \alpha \, \bm B_{\ell,\ell+1}  \big]
	\label{eq:dBPP_LDRL}
\end{align}

We determine the fixed point of the feedback weights by taking the expectation value over many noise samples for fixed inputs and weights,
\begin{align}
	0 &\stackrel{!}{=}\E \big[ \dot{\bm{B}}_{\ell,\ell+1} \big]_{\bm \xi} \\
	&= \E\Big[ \eta^\text{bw}_{\ell}  \big\{\bm \xi_\ell \, \big( \widehat{\bm r}_{\ell+1} \big)^T - \alpha \, \bm B_{\ell,\ell+1}  \big\} \Big]_{\bm \xi} \\
	\Rightarrow \quad &\E\big[\bm  B_{\ell,\ell+1} \big]_{\bm \xi} = \frac{1}{\alpha} \, \E \big[\bm \xi_\ell \, \big( \widehat{\bm r}_{\ell+1} \big)^T \big]_{\bm \xi}
\end{align}
Using Eq.~\eqref{eq:app_rP_noise_expanded}, the right hand side can be expanded,
\begin{align}
	\E\big[\bm  B_{\ell,\ell+1} \big]_{\bm \xi} &\approx \frac{1}{\alpha} \E\big[ \bm \xi_\ell \Big\{ \varphi'(\bm \ubreven_{\ell+1}) \,   \bm \xi_{\ell+1} + \varphi'(\bm \ubreven_{\ell+1}) \, \sum_{m=1}^{\ell} \big[ \prod_{n=m}^{\ell} \bm W_{n+1,n} \varphi'(\bm \ubreven_n) \big] \,   \bm \xi_m \Big\}^T \big]_{\bm \xi}
\end{align}
We now make use of the fact that the auto-covariance of Ornstein-Uhlenbeck noise decays exponentially in time,
\begin{align}
	\E \big[ \bm \xi_\ell (t+\Delta t) \; \bm \xi_k(t)^T \big]_{\bm \xi} = \id \, \delta_{k,\ell} \, \frac{\sigma^2}{2} \, e^{-|\Delta t|/\tauxi} \;,
	\label{eq:autocorr_noise}
\end{align}
where $\delta_{k,\ell}$ is the Kronecker delta, $\sigma^2$ the variance, and $\Delta t$ corresponds to the time needed for a noise sample to travel to the layer above and back.
As noise originating in different layers is uncorrelated, it quickly averages to zero, whereas the correlation of noise samples at different times generated at the same layer is non-zero.
\added{Note that PAL is compatible with different kinds of noise other than the Ornstein-Uhlenbeck type.
We only require the noise to have non-zero auto-correlation time, and noise from different sources (neurons) to show vanishing correlation.
This is also fulfilled e.g.~by pink noise, which is ubiquitous in biological systems~\cite{szendro2001bio} -- however, we argue that Ornstein-Uhlenbeck is a more natural model for noise in neurons due to the combination of stochastic inputs and synapses with slow dynamics.
Concretely, Refs.~\cite{ricciardi1979ornstein,petrovici2016stochastic} have shown that neurons with Poisson spike input and a membrane acting as a low-pass filter generate Ornstein-Uhlenbeck voltage dynamics.}

We thus have as our final result
\begin{align}
	\label{eq:app_BPP_fixed_point}
	\E\big[\bm  B_{\ell,\ell+1} \big]_{\bm \xi} &\approx \frac{1}{\alpha} \, \frac{\sigma^2}{2} \, e^{-|\Delta t|/\tauxi} \big\{\varphi'(\bm \ubreven_\ell) \, \big[ \bm W_{\ell+1,\ell} \big]^T  \varphi'(\bm \ubreven_{\ell+1}) \big\} \;.
\end{align}
The curly brackets show the alignment between feedback and feedforward weights.

\vspace{.25cm}

We now incorporate this result with the learning rule for bottom-up weights $\bm W_{\ell,\ell-1}$.
As discussed in Sec.~\ref{sec:learn_of_bw_weights}, error propagation mechanisms generally produce a layer-wise error $\bm e_\ell$ as a function of top-down synapses $\bm B_{\ell,\ell+1}$.

Models closely related to backpropagation~\cite{10.1162/089976603762552988,scellier2017equilibrium,haider2021latent} employ forward weight updates of the form
\begin{align}
	\label{eq:app_dWPP_1}
	\bm {\dot{W}}_{\ell,\ell-1} &= \eta^\text{fw}_\ell \, \bm e_\ell \, \bm r_{\ell-1}^T \nonumber\\
	&=\eta^\text{fw}_\ell \, \big\{\varphi'(\bm \ubreven_{\ell}) \, \bm B_{\ell,\ell+1} \,  \bm e_{\ell+1} \big\} \, \bm r_{\ell-1}^T\;,
\end{align}
\added{which applies to the model used to simulate Fig.~\ref{fig:eff_deep_nets} (Efficient credit assignment in deep networks).}
On the other hand, models where rates are propagated top-down~\cite{lee2015difference,sacramento2018dendritic,meulemans2020theoretical} can be generally described by
\begin{align}
	\label{eq:app_dWPP_2}
	\bm {\dot{W}}_{\ell,\ell-1} &= \eta^\text{fw}_\ell \, \big\{ \bm B_{\ell,\ell+1} \, \varphi'(\bm \ubreven_{\ell+1}) \, \bm e_{\ell+1} \big\} \, \bm r_{\ell-1}^T\;;
\end{align}
\added{e.g.,~the microcircuit model used in this work approximates this rule.}
We plug our result for learned top-down weights, Eq.~\eqref{eq:app_BPP_fixed_point}, into these update rules.
Because we have based our derivation on a general leaky-integrator model, our result in the form of $\E \big[ \bm  B_{\ell,\ell+1} \big] \propto \varphi'(\bm \ubreven_\ell) \, \big[ \bm W_{\ell+1,\ell} \big]^T  \varphi'(\bm \ubreven_{\ell+1}) \; \forall \;  \ell$ can be employed in any of these theories.
Plugging into either Eq.~\eqref{eq:app_dWPP_1} or~\eqref{eq:app_dWPP_2}, we see that the weight updates $\Delta \bm W_{\ell, \ell-1}$ align with those of a feed-forward network trained with backpropagation up to additional factors of derivatives.
Therefore, our algorithm can provide useful error signals which approximately align with backpropagation, improving on random feedback weights.

\added{Note that PAL is also able to learn useful backprojections in absence of a teaching signal, facilitating efficient learning once an instructive signal is (re-)introduced.
In particular, top-down weights do not decay to zero if forward weights are kept fixed, even though the weight decay term $\bm {\bm \dot{B}}_{\ell,\ell+1} \propto - \alpha \, \bm B_{\ell,\ell+1}$ might suggest so.
The reason for this is that the expectation value of top-down weights, Eq.~\eqref{eq:app_BPP_fixed_point}, are formed from a balance between noise and the contribution due to the regularizer.}

\subsection{Dendritic cortical microcircuits}
\label{app:Dendritic_cortical_microcircuits}

We now describe learning through minimization of the dendritic error as proposed in Ref.~\cite{sacramento2018dendritic}.
In this model, weights are adapted using local dendritic plasticity rules in the form $\bm{\dot{W}} = \eta \, [\varphi(\bm u) - \varphi(\bm v) ] \, \bm r^T $, where $\bm W$ represents lateral or feed-forward weights, $\eta$ is a learning rate, $\bm u$ and $\bm v$ denote different compartmental voltages and $\bm r$ the pre-synaptic rate~\cite{URBANCZIK2014521,10.3389/fncir.2018.00053}.
In order to adhere to the principle of bio-plausibility, the microcircuit model uses rules such that $\bm u$ corresponds to the soma of a given neuron and $\bm v$ to the corresponding compartment the synapse connects to.
The concrete form of all learning rules can be found in Supplementary Information.
They are designed such that the system settles in a specific state, where activity of the apical dendrite of hidden layer pyramidal cells represents an error useful for learning.

In absence of a teaching signal, pyramidal cells take the role of representation units, reflecting feed-forward activation.
They receive bottom-up information onto their basal dendrites and top-down activity in the distal apical dendrite, integrating both signals in the soma in accordance with observations of layer 2/3 pyramidal cells~\cite{jordan2020opposing}.
Pyramidal cells are described by a simplified three-compartment model with distinct basal, apical and somatic voltages~\cite{kording_supervised_2001,spruston_pyramidal_2008}.

The interneurons in this model are present in the hidden layers, and aim to represent a copy of the activation of pyramidal cells in the layer above.
Across layers, populations of pyramidal neurons and interneurons are arranged such that the number of interneurons in the hidden layers matches that of pyramidal cells in the layer above.
Interneurons are modeled with two compartments, representing dendritic tree and soma.
They receive lateral input from pyramidal cells in the same layer, and project back laterally to the same neurons.

We denote quantities related to pyramidal cells with upper indices $^\text{P}$, while $^\text{I}$ marks interneurons.
The dynamics of the somatic membrane potentials of pyramidal cells $\bm u^\text{P}_\ell$ with $\ell=1, \,\ldots\, ,N-1$ are an instance of the general leaky-integrator equation~\eqref{eq:LI_base_model},
\begin{align}
\Cm\bm {\dot u}^\text{P}_\ell &= \gl \left(\bm \El - \bm u^\text{P}_\ell\right) + g^\bas \left(\bm v^\bas_\ell - \bm u^\text{P}_\ell\right) + g^\api \left(\bm { v}^\api_\ell +\bm \xi_\ell(t) - \bm u^\text{P}_\ell\right)\; , \label{eq:mc_dyn}
\end{align}
where Ornstein-Uhlenbeck noise $\bm \xi_\ell$ is modeled at all hidden layers, and $\bm \El$ denotes the leak potential.
We implement Latent Equilibrium into the microcircuit model as in Ref.~\cite{haider2021latent} by replacing all rates calculated from somatic potentials with rates obtained from the prospective voltage, $\bm \rPI_{\ell} = \varphi(\bm { u}^\text{P/I}_{\ell}) \mapsto \varphi(\bm {\breve u}^\text{P/I}_{\ell})$.
This affects all compartment potentials as well as synaptic plasticity rules.

We first describe learning using fixed random feedback connections.
Before supervised training, the system is run in absence of a teaching signal with random input~$\bm r_0(t)$.
As demonstrated in Ref.~\cite{sacramento2018dendritic}, the plasticity of $\bm \WIP_{\ell, \ell}$ and $\bm \BPI_{\ell, \ell}$ (see Fig.~\ref{fig:mc_setup} for notation) allow the system to settle in a \textit{self-predicting state}.
This is a system state described by matching voltages between interneurons and pyramidal cells, $\bm u^\text{I}_\ell = \bm u^\text{P}_{\ell+1}$, together with zero apical voltages~$\bm v^\text{api}_\ell$.
This second condition is achieved in the hidden layers by learning lateral weights such that top-down and lateral activity cancel, i.e.~$\bm \BPI_{\ell, \ell} \, \bm \rI_\ell = - \bm \BPP_{\ell, \ell+1} \, \bm \rP_{\ell+1}$.

We now turn on an instructive signal by clamping the apical compartment of top layer pyramidal cells to the target voltage $\bm u^\text{tgt}$.
The somata of these neurons integrate the target signal with the bottom-up input and propagate it top-down to the hidden layers.
In the limit of small conductances (weak nudging), the pyramidal neurons are now slightly nudged towards the target, while the interneurons in the hidden layers do not observe the teaching signal; 
therefore, they represent the activity which pyramidal cells in the layer above would have if there were no target.
As the apical dendrite calculates the difference between top-down and lateral activity, it now encodes the error signal passed down to the hidden layers.
This is how cortical microcircuits with dendritic error encoding assign credit to hidden layers neurons.
In fact, the microcircuit model implements difference target propagation~\cite{lee2015difference} in a dynamical system with recurrency, with $\bm v_\ell^\text{api} = \bm \BPP_{\ell,\ell+1} ( \bm \rP_{\ell+1} - \bm \rI_\ell)$ representing the backprojected difference target $g_\ell(\hat{\bm h}_{\ell+1}) - g_\ell({\bm h}_{\ell+1})$.
\added{Note that the framework of DTP also includes the learning of top-down weights such that the system's cost is minimized according to Gauss-Newton optimization with batch size 1~\cite{meulemans2020theoretical}; this effect however is independent of the error propagation scheme and thus not relevant for this discussion.}

As shown in Ref.~\cite{sacramento2018dendritic}, the weight updates in this model approximate those of error backpropagation in the limit of weak nudging (small top-down conductances):
\begin{align}
	\label{eq:mc_dWPP_derivative}
	\Delta \bm \WPP_{\ell,\ell-1} &\propto \lambda^{N-\ell+1} \, \varphi' (\bm \vbashat_{\ell})  \, \big[ \prod_{k=\ell}^{N-1} \bm \BPP_{k, k+1} \, \varphi'( \bm \vbashat_{k+1} ) \big] \, \bm e_N \, \big(\bm \rP_{\ell-1} \big)^T\,,
\end{align}
where $\bm \vbashat_{\ell} \coloneqq \frac{\gbas}{\gl + \gbas + \gapi} \bm  v^\text{bas}_{\ell}$ denotes the conductance-weighted feed-forward input to each pyramidal cell, $\bm e_N \coloneqq \bm u^\text{tgt} -\bm \vbashat_{N} $ the output layer error, and a small parameter $\lambda$, which regulates the amount of top-down nudging.

PAL is implemented naturally by the inclusion of noise and our learning rule~\eqref{eq:IL_dBPP_LDRL}.
Contrary to simulations with fixed top-down weights~\cite{sacramento2018dendritic,haider2021latent}, the inclusion of PAL also requires dynamical lateral weights from interneurons to pyramidal cells.
For efficient learning, a tight balance between learning rates needs to be kept:
lateral weights $\bm \WIP_{\ell,\ell}$ need to adapt quickly to any changes of feed-forward weights $\bm \WPP_{\ell+1,\ell}$,
while top-down weights $\bm \BPP_{\ell,\ell+1}$ need to adapt to changing forward weights quickly;
on the other hand, lateral weights $\bm \BPI_{\ell,\ell}$ from interneurons to pyramidal cells need to adapt quickly to changing top-down weights, such that no spurious error occurs.
The precise order of weights updates is $|\Delta \bm \WPP_{\ell+1,\ell} | < |\Delta \bm \BPP_{\ell,\ell+1}| < |\Delta \bm \BPI_{\ell,\ell}| \lesssim |\Delta \bm \WIP_{\ell,\ell}|$.

We comment on the ability of cortical microcircuits to assign credit over hierarchies with multiple hidden layers.
While Eq.~\ref{eq:mc_dWPP_derivative} implies that in theory, tasks can be learned successfully using many hidden layers, the derivation assumes perfect cancellation of the top-down signal with the interneuron activity, such that only the error signal is encoded in the apical dendrite.
This requires a perfect self-predicting state, which is unattainable in practice unless learning is phased (learning phases interleaved with phases where no target is present and the self-predicting state is reestablished).
As the error signal scales with the small nudging strength~$\lambda$, early layers in the network receive instructive signals which are weaker by orders of magnitude, further complicating realistic evaluations of the model.
For this reason, we restrict our simulations showing credit assignment in cortical microcircuits to a single hidden layer (see Sections~\ref{sec:ts} \& \ref{sec:classification}).
We stress that this is a caveat of the microcirucit model, and not of PAL -- see our results in credit assignment in deep networks, Sec.~\ref{sec:deep_nets}.

\subsection{Simulation details}

We stress that all simulations are performed with fully recurrent dynamics described by Eqs.~\eqref{eq:LI_base_model} and~\eqref{eq:mc_dyn},
differentiating our work from similar studies where the dynamics are replaced by steady-state approximations and the recurrency is implicitly removed by calculating separate forward and backward passes~\cite{sacramento2018dendritic,greedy2022single}.
All simulation parameters are given in Supplementary Information.

In all experiments, dynamics are simulated in discrete time steps of length $dt$ using the Euler-Maruyama method~\cite{sarkka_solin_2019}.
Input and targets are passed as data streams in the form of vectors presented for $\Tpres = 100 \, dt$ without any kind of filtering or pre-processing.
In all simulations, voltage and weight updates (where non-zero) are applied at all time steps.
All layers are fully connected throughout all experiments.

Microcircuits are simulated by defining effective voltages $\bm u_\text{eff}$ as described in Ref.~\cite{haider2021latent}: we rewrite all dynamical equations in the form $C_m \bm{\dot{u}}  = \frac{1}{\taueff}(\bm u_\text{eff} - \bm u)$, where $\bm u $ denotes the pyramidal or interneuron somatic potential.
Models are initialized in the self-predicting state defined (see Supplement, and Ref.~\cite{sacramento2018dendritic}).
Before training, we allow the voltages to equilibrate during a brief settling phase (several~$dt$).
Activation functions are the same throughout all layers, including the output layer.
Targets are provided as a target voltage $\bm u^\text{tgt}$ at the output layer.

For the PAL implementations, we calculate Ornstein-Uhlenbeck noise by sampling white noise~$\bm w\sim \mathcal{N}(0,1)$ and low-pass filtering with time constant $\tauxi$, that is,
$ \bm \xi_\ell[t+dt] =  \bm \xi_\ell[t] +  \frac{1}{\tauxi}  (\sqrt{\tauxi \, dt} \, \sigma_\ell \, \bm w - dt \, \bm \xi_\ell[t])$.
High-pass filtered rates $\widehat{\bm r}_{\ell}$ are calculated with respect to the time constant $\tauhp$ through $\frac{d{\widehat{\bm r}}^\text{P}_\ell}{dt} =  \frac{d \bm r^\text{P}_\ell}{dt}  - \frac{{\widehat{\bm r}}^\text{P}_\ell}{\tauhp}$.
Forward weight updates low-pass filtered with $\taulo$ before application.

For pseudocode, all parameters and architecture details, see Supplementary Information.
Due to the intricate interplay of parameters specific to PAL ($\eta^\text{bw}, \sigma_\ell, \alpha, \taulo, \tauhp, \tauxi$) with the large number of parameters which the dendritic microcircuit framework already carries, we restricted our parameter search to the set of all learning rates and some of the PAL parameters, $\{\eta^\text{fw}, \eta^\text{bw}, \eta^\text{IP}, \eta^\text{PI}, \sigma_\ell, \alpha, \taulo\}$.
Simulations were optimized using these parameters by hand, ensuring the order of weight updates to comply without our findings of Sec.~\ref{app:Dendritic_cortical_microcircuits}.
The remaining parameters were set according to our assumptions about PAL, or, in the case of conductances, inherited from Ref.~\cite{haider2021latent}.

\subsubsection{Phaseless backwards weight alignment}

We simulate a microcircuit network of size [5-20-10-20-5] with sigmoid activation.
Linear and non-linear regimes are simulated by choosing bottom-up weights as
$\bm \WPP \sim \mathcal{U}[-1,1]$ and $ \sim \mathcal{U}[-5,5]$, respectively.
Forward weights are fixed, while top-down and lateral (inter- to pyramidal neuron) weights are learned. 

During evaluation against BP (right column in Fig.~\ref{fig:bw_learning}), we set the lateral weights during to the exact self-predicting state, in order to observe an error signal in earlier hidden layers.
Note that the weight updates in the microcircuit model do not exactly represent those of an ANN.
This is due to an additional factor $\varphi'(\bm \ubreveP_N)$ as well as the fact that top-down nudging influences the somatic activity (see Supplementary Information for details).
These factors introduce a misalignment unrelated to the performance of PAL, as can be seen through the comparison of updates in the non-linear case~(Fig.~\ref{fig:bw_learning} (f));
in particular, already in the output layer ($\ell = 4$), a misalignment of $\sim 20^\circ$ can be observed, and hence even perfectly learned backprojections are likely to observe increased misalignment.
Therefore, it is to be expected that alignment is improved further in theories of error propagation which relate more closely to backpropagation.

In order to illustrate the advantages of prospective coding (Fig.~\ref{fig:PAL_needs_LE}), we re-ran the experiments of Fig.~\ref{fig:bw_learning}~(a,b,c) with time constants $\{\Tpres, \tauhp, \tauxi\}$ rescaled by 10 and 100, while rescaling all learning rates by the inverse.
Runs without prospective coding used the rate output $\bm r_\ell = \varphi(\bm u_\ell)$ instead of $\varphi(\bm \ubreve_\ell)$.

\subsubsection{Classification experiments}

The Yin-yang and MNIST tasks were solved using microcircuit networks of size [4-30-3] and [784-100-10], respectively, with sigmoid activation.
All weights (including lateral) were trained with fully recurrent dynamics.
For the experiments shown in this section we use the GPU enhanced Neuronal Network simulation environment (GeNN)~\cite{yavuz2016genn,knight2021pygenn}.
Natively, GeNN supports the simulation of spiking neural networks, but the possibility to add custom neuron and synapse models to the already provided ones makes the implementation of rate-based models such as the dendritic microcircuit possible.
The simulation of the dendritic microcircuits benefited greatly from the GPU support provided by GeNN which allowed us to perform the experiments shown in this section within practically feasible simulation times.

As shown in Fig.~\ref{fig:teacher_student} (c,d,e,f), PAL is able to outperform FA in benchmark tasks.
However, this may be the case simply due to the inclusion of noise on neuronal dynamics.
In order to demonstrate that top-down learning with PAL is really required for increased performance,
we have simulated PAL with top-down learning turned off, which is equivalent to a network trained with FA with noise.
The results are shown in Fig.~\ref{fig:FA_with_noise} in Supplementary Information and show that indeed, top-down alignment facilitated by PAL is required for increased performance.

\subsubsection{Efficient credit assignment in deep networks}

For these experiments, we simulated the general leaky integrator model, Eq.~\eqref{eq:LI_base_model}.
As previously, this model was simulated in discrete time steps, with images presented for $\Tpres = 100~\text{dt}$.
Voltages were updated continuously, and weight updates are applied at all steps.
The output layer error was defined as $\bm e_N = \bm u^\text{tgt} - \bm \ubreve_{N}$\added{, and hidden layer errors are calculated via $\bm e_\ell = \varphi' \cdot \bm B_{\ell,\ell+1} [\bm {\breve{u}}_{\ell+1} - \bm W_{\ell+1,\ell} \, \bm {{r}}_{\ell}$]}.
We have also included a bias term for each neuron.
PAL was implemented by adding Ornstein-Uhlenbeck noise $\bm \xi_\ell$ to each hidden layer neuron, calculating the high-pass filtered rate $\widehat{\bm r}_{\ell+1}$ and updating top-down weights with Eq.~\eqref{eq:IL_dBPP_LDRL}.

Network size of the autoencoder was [784-200-2-200-784], with activations [tanh, linear, tanh, linear].
We define as latent space activity the output of the two neurons in the central hidden layer.
After every epoch of training the LI model, we trained a linear classifier on the MNIST train set and show the accuracy of the linear classifier on the test set.

\added{For the CIFAR-10 experiments, we built on the continuous-time variant of LeNet-5 of Ref.~\cite{haider2021latent}.}
\added{In order to achieve our results on these time-consuming simulations, we have made a simplification: top-down signals (errors) are not added to the somata, such that the full soma dynamics are}
\begin{align}
	\tau \bm {\dot u}_\ell = - \bm u_\ell + \bm b_\ell + \bm W_{\ell,\ell-1} \bm r_{\ell-1} + \bm \xi_\ell \,.
\end{align}
\added{This weakens the recurrency of all signals in the network, making it easier to train with all algorithms.}
\added{As a consequence, error signals are passed top-down directly as $\bm e_\ell = \varphi' \cdot \bm B_{\ell,\ell+1} \bm e_{\ell+1}$.}

The network consists of four layers: Conv2d($(5\times 5) \times 20$) with sigmoid and MaxPool(2) activation, Conv2d($(5\times 5) \times 50$) with sigmoid and MaxPool(2) activation, a fully-connected projection layer of 500 neurons and a linear output layer with 10 neurons.
With PAL, we train the top-down weights from the second Conv2d layer to the projection layer and from the projection to the output layer, as the remaining bottom-up connection (Conv2d to Conv2d) is a kernel, for which PAL is not defined.
For the second layer, noise is added after the MaxPool(2) activation instead of on top of the somata, as this increases convergence time of top-down weights.
Validation and test accuracies are determined without noise, in order to enable a fair comparison with other methods.

In these experiments, hyperparameter optimization was carried out by a grid search over $\eta^\text{fw}$, while the remaining parameters were optimized manually.
See Tab.~\ref{tab:sim_values_DL}~in Supplement for parameters.\\

\section*{Data availability statement}\label{sec:data_avail}
The datasets analyzed during the current study are publicly available in the following repositories:\\
\url{https://github.com/lkriener/yin_yang_data_set} (Yin-Yang dataset~\cite{kriener2021yin}), \url{http://yann.lecun.com/exdb/mnist/} (MNIST dataset~\cite{lecun1998gradient}), 
\url{https://www.cs.toronto.edu/~kriz/cifar.html} (CIFAR-10~\cite{krizhevsky2009learning}).

\section*{Code availability}\label{sec:code_avail}
All code is made available under \url{https://doi.org/10.5281/zenodo.10401883}.

\section*{Acknowledgment}\label{sec:ack}
We wish to thank Jakob Jordan, Alexander Meulemans and Jo\~{a}o Sacramento for valuable discussions.
We gratefully acknowledge funding from the European Union under grant agreements 604102, 720270, 785907, 945539 (HBP) and the Manfred St{\"a}rk Foundation.
Additionally, our work has greatly benefited from access to the Fenix Infrastructure resources, which are partially funded from the European Union's Horizon 2020 research and innovation programme through the ICEI project under the grant agreement No.~800858.
This includes access to Piz Daint at the Swiss National Supercomputing Centre, Switzerland.
Further calculations were performed on UBELIX, the HPC cluster at the University of Bern.

\section*{Author contributions}
K.M.~derived, with contributions by L.K.~and M.A.P., the phaseless alignment learning (PAL)
algorithm. K.M.~and L.K.~adapted the dendritic microcircuit model to include PAL for learning
the feedback weights. G.G.~and T.N.~developed a dendritic microcircuit module for the GeNN
simulator. L.K.~added the latent equilibrium and PAL mechanisms to the module. K.M.~and
L.K.~performed the simulation experiments.
\added{I.J.~and K.M.~worked on scaling the algorithm to a larger benchmark during the revision process.}
The manuscript was mainly written by K.M., aided by L.K.~and M.A.P.
M.A.P.~and W.S.~provided supervision and funding to this project.

\section*{Competing interests}
The authors declare no competing interests.

\newpage

\FloatBarrier
\printbibliography{}
\addcontentsline{toc}{section}{References}

\FloatBarrier
\pagebreak

\clearpage
\appendix

\renewcommand{\figurename}{Supplementary Figure}
\setcounter{figure}{0}

\section{Additional information on PAL}

In this Supplement, we give more detail on the derivation and application of PAL (App.~\ref{app:align_bw}) and the microcircuit implementation (App.~\ref{app:app_error_BP_MC}) used to perform the simulations.

As in the main text, bold lowercase (uppercase) variables $\bm x$ ($\bm X$) denote vectors (matrices).
The partial derivative of the activation given by $\bm r_\ell = \varphi (\bm {\breve{u}})$ is denoted by
$\varphi'(\bm {\breve{u}})$, which is a diagonal matrix with $\mu$-th entry $\frac{\partial r_\mu}{\partial  {\breve{u}_\mu} }$.

\subsection{Derivation of PAL}

\label{app:align_bw}

We point out how and why our alignment loss $\mathcal{L}^\text{PAL}_\ell$, defined in Eq.~\eqref{eq:L_PAL}, differs from the reconstruction loss $\widehat{\mathcal{L}}$ introduced in Ref.~\cite{ernoult2022towards}.
Using the notation of this manuscript, this can be expressed as
\begin{align}
	    \widehat{\mathcal{L}}_{\bm B_{\ell,\ell+1}}^\ell &\equalhat -\bm  \xi_\ell^T \, \bm B_{\ell,\ell+1} \big\lbrace \bm r^{\bm \xi_\ell}_{\ell+1} - \varphi(\bm \ubreven_{\ell+1}) \big\rbrace + \left\| \bm B_{\ell,\ell+1} \big\lbrace \bm r^{\bm \xi_{\ell+1}}_{\ell+1} - \varphi(\bm \ubreven_{\ell+1}) \big\rbrace \right\|^2\;,
\end{align}
where $\bm r^{\bm \xi_\ell}_{\ell+1}$ is the rate which comprising data signal and noise from layer $\ell$ only, while the second term generated from $\bm \ubreven_\ell$ contains no noise signal.
The regularizer requires a separate phase, where noise is injected only into layer $\ell+1$ and backpropagated to layer $\ell$.

Training feedback weights by gradient descent on this alignment loss represents a case closely related to PAL -- for fixed input and forward weights, $\bm B_{\ell,\ell+1}$ converges such that the Jacobians matrices align, $\bm B_{\ell,\ell+1}\,  \varphi'(\bm \ubreven_{\ell+1}) \, \| \, [\varphi'(\bm \ubreven_{\ell}) \, \bm W_{\ell+1,\ell} ]^T$.
Inserting this result into the difference target propagation rule of Eq.~\eqref{eq:app_dWPP_2}, we see that the update reproduces exact backpropagation with linear activation function on the output layer.

An analogous implementation of gradient descent on $\widehat{\mathcal{L}}$ in our setup requires making use of the noise in the output layer, and using a different kind of regularizer, i.e.~$- \alpha  \, \| \bm B_{\ell, \ell+1} \widehat{\bm r}_{\ell+1} \|^2$ instead of weight decay.
Unfortunately, this regularizer contains non-zero correlations of all noise signals up to layer $\ell+1$, and not only auto-correlations of $\bm \xi_\ell$.
Therefore, gradient descent on a difference reconstruction loss with this regularizer does not lead to useful top-down weights, as a particular weight $\bm B_{\ell, \ell+1}$ receives contributions proportional to all weights $\bm W_{k+1,k}$ for $k=1\,\ldots\,\ell$.
The central reason causing this issue is that our system learns to adapt all feedback weights simultaneously, which requires considering noise in all layers at all times.
Contrary to this, in Ref.~\cite{ernoult2022towards}, feedback weights are trained sequentially with two separate phases of noise injections in different layers.
We have therefore designed Eq.~\eqref{eq:dBPP_LDRL} as a heuristic approximation to the optimal update rule%
, while achieving full always-on plasticity in our system.

\vspace{.25cm}

\begin{figure}[t]
	\centering
	\includegraphics[width=.7\textwidth]{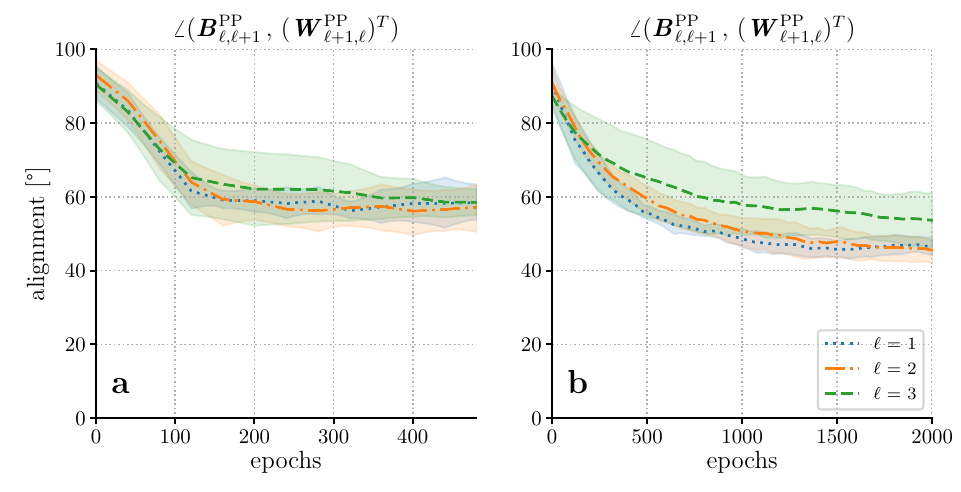}
	\caption{
		\textbf{Alternative regularizer with derivative shows further improvement in alignment.}
		We reproduce the experiment in Fig.~\ref{fig:bw_learning} (e) using the same parameters: microcircuits learning to adapt backwards weights with PAL using \textbf{(a)} the standard weight decay regularizer and \textbf{(b)} the derivative-dependent regularizer of Eq.~\eqref{eq:varphi_regularizer}.
	}
	\label{fig:bw_alignment_varphi}
\end{figure}

Alternatively, the problem of superfluous derivatives $\varphi'$ can be addressed if the derivative of the activation w.r.t.~to the potential is available at the synapse.
Given this information, the weight updates can be defined as
\begin{align}
	\bm \dot{\bm{B}}_{\ell,\ell+1} &= \eta^\text{bw}_{\ell}  \big[\bm \xi_\ell \; \big( \widehat{\bm r}_{\ell+1} \big)^T - \alpha \, \varphi'(\bm {\breve{u}}_\ell) \bm B_{\ell,\ell+1} \varphi'(\bm {\breve{u}}_{\ell+1}) \big] \;.
	\label{eq:varphi_regularizer}
\end{align}
Note that the derivatives $\varphi'(\bm {\breve{u}}_\ell)$ are a function of the full somatic potential, comprising data as well as noise.
Using the fact that correlations between $\bm \xi_\ell$ and $\bm \xi_{\ell+1}$ cancel to zero, we obtain that the new expectation value of top-down weights to first order,
\begin{align}
	\E\big[\varphi'(\bm \ubreven_\ell) \bm  B_{\ell,\ell+1} \varphi'(\bm \ubreven_{\ell+1}) \big]_{\bm \xi} &\propto\varphi'(\bm \ubreven_\ell) [\bm  W_{\ell+1,\ell}]^T \varphi'(\bm \ubreven_{\ell+1})  \;.
\end{align}
We have tested this alternative regularizer in the relevant regime of in non-linear activation.
As shown in Fig.~\ref{fig:bw_alignment_varphi}, it is able to improve alignment of $\bm \BPP_{\ell, \ell+1}$ with $[\bm \WPP_{\ell+1,\ell}]^T$ by about~$10^\circ$ in this example (note that we have used the same parameters as in the PAL setup. With appropriate hyperparameter search, convergence time and final alignment may be improved).
However, whether a bio-plausible synapse can calculate the derivative $\varphi' = \frac{\partial \varphi}{\partial \bm \ubreve}$ is not clear.
Given our requirement that all computations can be implemented with simple physical components, we have opted for the weight decay regularizer as defined in Eq.~\eqref{eq:L_PAL}.

\subsection{Error propagation in microcircuits}
\label{app:app_error_BP_MC}

In this section, we explain in detail how the cortical microcircuit is able to propagate meaningful targets, and align its feedback weights using PAL in order to efficiently minimize the difference between its output and a teaching signal.

We briefly review the general microcircuit setup defined by Ref.~\cite{sacramento2018dendritic}.
In this model, the different neuron populations are each selected to play a distinct role.
Each hidden layer is composed of a population of pyramidal neurons and interneurons, where the number of interneurons in a given layer matches the number of pyramidal cells in the layer above.
The neurons form a network defined by connections as shown in Fig.~\ref{fig:mc_setup}.

As in Ref.~\cite{sacramento2018dendritic}, we define the following coupled differential equations to govern the voltage dynamics of pyramidal cells ($\bm u^\text{P}_\ell$) and interneurons ($\bm u^\text{I}_\ell$) in a network with layers $\ell = 1\, \ldots \, N$:
\begin{align}
	\Cm\bm {\dot u}^\text{P}_\ell &= \gl \left(\bm \El - \bm u^\text{P}_\ell\right) + g^\bas \left(\bm v^\bas_\ell - \bm u^\text{P}_\ell\right) + g^\api \left(\bm { v}^\api_\ell +\bm \xi_\ell(t) - \bm u^\text{P}_\ell\right)\; \quad \forall \ell \neq N , \label{eq:dyn1}  \\
	\Cm\bm {\dot u}^\text{P}_N &= \gl \left(\bm \El - \bm u^\text{P}_N\right) + g^\bas \left(\bm v^\bas_N - \bm u^\text{P}_N\right) + \bm i^\text{nudge,tgt} \;,\label{eq:dyn2} \\
	\Cm\bm {\dot u}_\ell^\text{I} &= \gl \left(\bm \El - \bm u_\ell^\text{I}\right) + g^\den \left(\bm v^\den - \bm u_\ell^\text{I}\right) +  \bm i^\text{nudge,I} \; .\label{eq:dyn3}
\end{align}
Here, $\bm \xi_\ell$ denotes the noise modeled in all hidden layers.
Compartment voltages are induced instantaneously by the respective input rates and synaptic weight,
\begin{align}
	\bm v^\bas_\ell &= \bm \WPP_{\ell, \ell-1} \varphi\left( \bm {\breve u^\text{P}}_{\ell-1}\right) \label{eq:dendr_volts1} \; , \\
	\bm v^\api_\ell &= \bm \BPP_{\ell, \ell+1} \varphi\left( \bm {\breve u^\text{P}}_{\ell+1}\right) + \bm \BPI_{\ell, \ell} \; \varphi\left(\bm {\breve u^\text{I}}_\ell \right) \; , \\
	\bm v^\den_\ell &= \bm \WIP_{\ell, \ell} \varphi\left( \bm {\breve u}^\text{P}_{\ell} \right) \; . \label{eq:vden}
\end{align}
The nudging currents for the interneurons are $\bm i^\text{nudge,I} = g^\text{nudge,I}( \bm \ubreveP_{\ell+1} - \bm \ubreveI_\ell )$, and the output layer pyramidal neurons receive a weak instructive signal via $\bm i^\text{nudge,tgt} = g^\text{nudge,tgt}( \bm u^\text{tgt} - \bm \ubreveP_N )$.

In the base microcircuit model of Ref.~\cite{sacramento2018dendritic} augmented with prospective coding~\cite{haider2021latent}, synaptic plasticity of forward and lateral weights is defined as
\begin{align}
	\bm {\dot W}^\text{PP}_{\ell, \ell-1} &= \eta^\text{fw}_{\ell}\left[\varphi\left(\bm {\breve u^\text{P}}_\ell\right) - \varphi\left(\frac{g^\bas}{\gl + g^\bas + g^\api} \bm v^\bas_\ell \right) \right] \varphi\left(\bm {\breve u}^\text{P}_{\ell-1}\right)^T \label{eq:mc_wdot1} \quad  \forall \ell \neq N \; , \\
	\bm {\dot W}^\text{PP}_{N, N-1} &= \eta^\text{fw}_{N}\left[\varphi\left(\bm {\breve u^\text{P}}_N\right) - \varphi\left(\frac{g^\bas}{\gl + g^\bas } \bm v^\bas_N \right) \right] \varphi\left(\bm {\breve u}^\text{P}_{N-1}\right)^T \label{eq:mc_wdot2} \; , \\
	\bm {\dot W}^\text{IP}_{\ell, \ell} &= \eta^\text{IP}\left[\varphi\left(\bm {\breve u}_\ell^\text{I}\right) - \varphi\left(\frac{g^\den}{\gl + g^\den} \bm v^\den_\ell \right) \right] \varphi\left(\bm {\breve u}^\text{P}_{\ell}\right)^T \; ,\label{eq:mc_wdot3} \\
	\bm \dot{\bm{B}}^\text{PI}_{\ell, \ell} &= \eta^\text{PI}_{\ell}\left[- \bm v^\api_\ell \right] \varphi\left(\bm {\breve u}^\text{I}_\ell\right)^T\,, \label{eq:mc_wdot4}
\end{align}
while top-down weights $\bm \BPP_{\ell, \ell+1}$ are fixed.

Before deriving the relevant analytical expressions, we briefly explain how the design of the circuitry leads to well-defined error propagation.
In absence of a teaching signal, and if the microcircuit has settled into its self-predicting state, the interneuron activity in each layer represents an exact copy of the pyramidal neurons in the layer above.
Interneurons project laterally onto the apical dendrites of pyramidal cells in the same layer;
these apical dendrites also receive input from pyramidal cells in the layer above.
In the self-predicting state, these activities are subtracted from each other;
as they are exactly the same, the inputs cancel, and the apical compartment voltage is zero.

We now introduce a weak nudging signal towards the correct voltage at the output layer.
To first order in expansion parameters, the interneurons still represent the pyramidal neurons in the layer above \textit{in absence of a teaching signal}.
The activity of the pyramidal cell in the layer above now however additionally contains the error signal.
Therefore, the difference in activity calculated at the apical dendrite in a given layer also represents an error.
Starting from the penultimate layer, this argument extends successively to the apical compartment voltages in all hidden layers.
Consequently, the apical compartments represent errors useful for learning, and these errors are backpropagated.

In order to prove the above statements, we reconsider the dynamics defined by Eqs.~\eqref{eq:dyn1}--\eqref{eq:dyn3} without noise, and learning rules~\eqref{eq:mc_wdot1}--\eqref{eq:mc_wdot4}.
Before performing supervised training, the system must settle in a self-predicting state.
This is achieved by presenting input sequences while clamping the target voltage to the prospective voltage, $\bm u^\text{tgt} = \bm \ubreveP_N$, and evolving the system while keeping the bottom-up weights $\bm \WPP_{\ell+1,\ell}$ and top-down weights $\bm \BPP_{\ell,\ell+1}$ fixed.
The dynamics of the lateral weights, $\bm {\dot W}^\text{IP}_{\ell, \ell} $ and $ \bm \dot{\bm{B}}^\text{PI}_{\ell, \ell} \,$, are designed to drive the respective weights to the self-predicting state and are required to work in conjunction.

The lateral connections from interneurons to pyramidal cells $\bm \BPI_{\ell, \ell} \,$ are driven by gradient descent on the mismatch energy $\| \bm \vapi_\ell \|^2$ defined by the apical compartment potential.
For fixed top-down synapses, the dynamics of Eq.~\eqref{eq:mc_wdot4} settle such that $\bm \vapi_\ell$ is (approximately) zero for all inputs.

Through the dynamics of Eq.~\eqref{eq:mc_wdot3}, the weights  $\bm {W}^\text{IP}_{\ell, \ell} $ are adapted to minimize the difference between the dendritic potential of the interneurons and the voltage in the basal compartment of pyramidal cells in the layer above.
This can be seen by expanding the learning rule in $g^\text{nudge,I} \ll \gl + \gden$,
\begin{align}
    \bm {\dot W}^\text{IP}_{\ell, \ell}  & \propto \varphi\left(\bm {\breve u}_\ell^\text{I}\right) - \varphi \Big(\frac{g^\den}{\gl + g^\den} \bm v^\den_\ell \Big) \\
    & = \varphi\Big( \frac{g^\den \bm v^\den_\ell + g^\text{nudge,I}  \bm \ubreveP_{\ell+1}}{\gl + g^\den + g^\text{nudge,I} }   \Big) - \varphi\Big(\frac{g^\den}{\gl + g^\den} \bm v^\den_\ell \Big) \\
    &\approx \varphi'\Big(\frac{g^\den}{\gl + g^\den} \bm v^\den_\ell\Big) \; \frac{g^\text{nudge,I}}{\gl + g^\den } \; \big[ \bm {\breve u}_{\ell+1}^\text{P} - \frac{g^\den}{\gl + g^\den} \bm v^\den_\ell \big]\;,
\end{align}
where in the first step, we have replaced the prospective interneuron voltage with the potentials which induce it, $\bm \ubreveI_\ell = \frac{g^\den \bm v^\den_\ell + g^\text{nudge,I}  \bm \ubreveP_{\ell+1}}{\gl + g^\den + g^\text{nudge,I} } $ given by Eq.~\eqref{eq:dyn3}.
In the second step, we expand in weak nudging of the interneuron.

In conjunction with the minimization of the apical potential in all layers through $ \bm \dot{\bm{B}}^\text{PI}_{\ell, \ell} \,$, the prospective potential $\bm {\breve u}_{\ell+1}^\text{P}$ is fully determined by its basal input, $\bm {\breve u}_{\ell+1}^\text{P} = \frac{\gbas}{\gl + \gbas + \gapi} \bm  v^\text{bas}_{\ell+1}$ for $1 \leq \ell < N-1$ and  $\bm {\breve u}_{N}^\text{P} = \frac{\gbas}{\gl + \gbas} \bm  v^\text{bas}_{N}$ for the output layer.
Therefore, the synapses settle into a state which minimizes the difference between the basal voltage $ \bm  v^\text{bas}_{\ell+1}$ and the interneuron compartment $\bm v^\den_\ell$ (up to a factor defined by the conductances) for all input samples.

After the lateral weights have converged, the interneuron potentials are an exact copy of the pyramidal cells in the layer above, $\bm \ubreveI_\ell = \bm \ubreveP_{\ell+1}$;
this can be seen by plugging the steady state solution $g^\den \bm v^\den_\ell =  (\gl + g^\den) \, \bm {\breve u}_{\ell+1}^\text{P} $ into the expression of $\bm \ubreveI_\ell$ in terms of its compartmental voltages.

A particularly well-suited self-predicting state is defined by
\begin{align}
	\bm \BPI_{\ell,\ell} &= - \bm \BPP_{\ell,\ell+1} \nonumber\\
	\bm \WIP_{\ell,\ell} &= \frac{\gbas}{\gden}  \frac{\gl + \gden}{\gl + \gbas + \gapi}  \bm  \WPP_{\ell+1,\ell}
	\label{eq:app_the_SPS}
\end{align}
for hidden layers, and $\bm \WIP_{N-1,N-1} = \frac{\gbas}{\gden}  \frac{\gl + \gden}{\gl + \gbas}  \bm  \WPP_{N,N-1}$ for the lateral weights projecting to the interneurons in the final hidden layer.
This state has the advantage that the lateral weights form a self-predicting state independent of the input data (general solutions of $\bm {\dot W}^\text{IP}_{\ell, \ell} = 0 = \bm \dot{\bm{B}}^\text{PI}_{\ell, \ell} $  do not perform as well in practice, as stimulus switching often requires re-learning of lateral weights before apical compartments represent useful error signal).
In the simulations presented in this work, the networks are initialized in this specific self-predicting state.

\vspace{.25cm}

We now turn on a teaching signal $\bm u^\text{tgt}$.
The new, nudged prospective state of the output neurons is $\bm \ubreveP_N = \frac{\gbas \bm v^\text{bas}_N + g^\text{nudge,tgt} \bm u^\text{tgt}}{\gl + \gbas + g^\text{nudge,tgt}}$.
Inserting this state into Eq.~\eqref{eq:mc_wdot2} and expanding the somatic potential about the weighted basal input $\bm \vbashat_{N}$, we obtain
\begin{align}
	\Delta \bm \WPP_{N,N-1} \approx \varphi' (\bm \vbashat_{N}) \cdot \frac{g^\text{nudge,tgt}}{\gl+\gbas+g^\text{nudge,tgt}} \big[\bm u^\text{tgt} - \bm \vbashat_{N} \big] \; \big(\bm \rP_{N-1} \big)^T \;.
\end{align}
Here, we have defined $\bm \vbashat_{N} \coloneqq \frac{\gbas}{\gl + \gbas} \bm  v^\text{bas}_{N}$, and we have rewritten the bottom-up input from pyramidal neurons in the penultimate layer as a rate $\bm \rP_{N-1}$.
We can now identify the difference between target and bottom-up input as the output layer error, $\bm e_N \coloneqq \bm u^\text{tgt} - \bm \vbashat_{N}$, and we may thus write
\begin{align}
	\label{eq:app_dWPPN}
	\Delta \bm \WPP_{N,N-1} \propto \varphi' (\bm \vbashat_{N}) \, \bm e_N \, \big(\bm \rP_{N-1} \big)^T\;.
\end{align}
Written in this form, we have demonstrated that the update rule~\eqref{eq:mc_wdot2} implements error minimization on the output layer in the limit of weak nudging.
One can regard this as equivalent to training the output layer of a feed-forward network, evaluated at the weighted input $\bm \vbashat_N$.

We now aim to show that the error $\bm e_N$ is propagated backwards through the network, where the apical compartment voltages represent the local error within each layer.
Starting from the self-predicting state, the apical voltages are
\begin{align}
	\bm v^\api_{N-1} &= \bm \BPP_{N-1, N} \big[ \varphi\left( \bm {\breve u^\text{P}}_{N}\right) - \varphi\left(\bm {\breve u^\text{I}}_{N-1} \right) \big]\\
	&= \bm \BPP_{N-1, N} \big[ \varphi \Big( \frac{\gbas \, \bm v^\text{bas}_N + g^\text{nudge,tgt} \, \bm u^\text{tgt}}{\gl + \gbas + g^\text{nudge,tgt}} \Big) - \varphi\left( (1- \lambda^\text{I}) \, \bm \vbashat_{N} + \, \lambda^\text{I} \, \frac{\gbas \, \bm v^\text{bas}_N + g^\text{nudge,tgt} \, \bm u^\text{tgt}}{\gl + \gbas + g^\text{nudge,tgt}} \right) \big]\nonumber \\
        \bm v^\api_\ell &= \bm \BPP_{\ell, \ell+1} \big[ \varphi\left( \bm {\breve u^\text{P}}_{\ell+1}\right) - \varphi\left(\bm {\breve u^\text{I}}_\ell \right) \big]\\
         &= \bm \BPP_{\ell, \ell+1} \big[ \varphi\left( \bm \vbashat_{\ell+1} + \lambdaP \, \bm \vapi_{\ell+1} \right) - \varphi\left( \bm \vbashat_{\ell+1} + \lambda^\text{I} \, \lambdaP \, \bm \vapi_{\ell+1} \right) \big]\nonumber
\end{align}
with $\lambda^\text{I} \coloneqq \frac{g^\text{nudge,I}}{\gl + \gden + g^\text{nudge,I}}$, $\lambdaP \coloneqq \frac{\gapi}{\gl + \gbas + \gapi}$, and $\bm \vbashat_{\ell+1} \coloneqq \frac{\gbas}{\gl + \gbas + \gapi} \bm  v^\text{bas}_{\ell+1}$.
Let us first focus on the apical voltage in the penultimate layer.
We again make use of the assumption of weak nudging, and additionally require that the interneuron is only weakly nudged by the top-down input it receives, $\lambda^\text{I} \ll 1$.
The apical voltage takes the form
\begin{align}
	\label{eq:app_vapiN_BP}
	\bm v^\api_{N-1} &\approx \bm \BPP_{N-1, N} \, \varphi'\left( \bm \vbashat_{N} \right) \frac{g^\text{nudge,tgt}}{\gl+\gbas} \bm e_N \,.
\end{align}
In the same fashion, we expand the apical potentials in the layers below.
The first order in $\lambdaP \, \bm v^\text{api}_{\ell+1}$ and zeroth order in $\lambda^\text{I}$ yields
\begin{align}
	\label{eq:app_vapiell_BP}
        \bm v^\api_\ell &\approx \bm \BPP_{\ell, \ell+1} \, \varphi'\left( \bm \vbashat_{\ell+1} \right) \lambdaP \, \bm v^\text{api}_{\ell+1} \,.
\end{align}
Taken together, these two results show that the apical potentials represent errors which are successively propagated backwards through the network.

Finally, the last missing ingredient is how apical errors are used to update the forward weights.
Performing the same expansion in small apical voltages on Eq.~\eqref{eq:mc_wdot1}, we obtain
\begin{align}
	\label{eq:app_dWPP_expanded}
		\Delta \bm \WPP_{\ell,\ell-1} &\propto \lambdaP \varphi' (\bm \vbashat_{\ell})  \, \bm v^\api_\ell \, \big(\bm \rP_{\ell-1} \big)^T\,.
\end{align}

We now collect and summarize our findings.
Bottom-up weights $\bm \WPP_{\ell,\ell-1}$ are updated using the local error, represented by the apical voltage as $\bm \vapi_\ell$, multiplied with the bottom-up signal $\bm  \rP_{\ell-1}$.
Eqs.~\eqref{eq:app_vapiN_BP} and~\eqref{eq:app_vapiell_BP} show that these errors are backpropagated by multiplying with the derivative $\varphi'(\bm \vbashat_{\ell+1})$ and feedback weights $\bm \BPP_{\ell,\ell+1}$.
This learning scheme resembles that of feed-forward networks trained with feedback alignment (for fixed $\bm \BPP$), or backpropagation (if $\bm \BPP_{\ell,\ell+1} = (\bm \WPP_{\ell+1,\ell})^T$).
One marked difference to a feed-forward network is the emergence of the derivative $\varphi' (\bm \vbashat_{N})$ in the update rule to all bottom-up weights, see Eq.~\eqref{eq:app_dWPPN} and~\eqref{eq:app_vapiN_BP}.
As we have defined the error $\bm e_N$ on the voltage level, one may expect there to be no such derivative, as one finds in the corresponding case of a feed-forward networks trained with backpropagation with linear activation functions on the output layer.
This additional factor signals a fundamental difference in architecture between backpropagation and difference target propagation, of which dendritic cortical microcircuits are an implementation.
In difference target propagation, targets are constructed locally from backpropagated rates -- i.e.~a target potential $\bm u^\text{tgt}$ is converted into a rate before it can be passed to a lower layer.
In contrast, in backpropagation, top-down signals are given by errors, bypassing the activation function in the upper layer and instead directly transporting potential differences to hidden layers.

We now incorporate this result with PAL.
As shown in Eq.~\eqref{eq:app_BPP_fixed_point} in App.~\ref{app:align_bw}, using PAL, the top-down weights converge to $\E \big[ \bm  B_{\ell,\ell+1} \big] \propto \varphi'(\bm \ubreven_\ell) \, \big[ \bm W_{\ell+1,\ell} \big]^T  \varphi'(\bm \ubreven_{\ell+1}) \; \forall \;  \ell$.
In the microcircuit model, forward weights are learned as derived in Eq.~\eqref{eq:app_dWPP_expanded};
in summary, we have found that
\begin{align}
	\label{eq:app_bw_dWPP}
	\Delta \bm \WPP_{\ell,\ell-1} &\propto \varphi' (\bm \vbashat_{\ell})  \, \big[ \prod_{k=\ell}^{N-1} \bm \BPP_{k, k+1} \, \varphi'( \bm \vbashat_{k+1} ) \big] \, \bm e_N \, \big(\bm \rP_{\ell-1} \big)^T\,.
\end{align}
In the limit of weak nudging and feedback, the noise-free potentials $\bm \ubrevePn_\ell$ are well approximated by the bottom-up input~$\bm \vbashat_\ell$, and our result for top-down weights takes the form $\E \big[ \bm  \BPP_{\ell,\ell+1} \big] \propto \varphi'(\bm \vbashat_\ell) \, \big[ \bm \WPP_{n+1,n} \big]^T  \varphi'(\bm \vbashat_{\ell+1}) \; \forall \;  \ell$.
Plugging this into Eq.~\eqref{eq:app_bw_dWPP}, we see that the weight updates $\Delta \bm \WPP_{\ell, \ell-1}$ align with those of a feed-forward network trained with backpropagation up to additional factors of derivatives.
Therefore, our algorithm can provide useful error signals which approximately align with backpropagation, improving on random feedback weights.

\subsection{Local alignment is compatible with approximate Gauss Newton-target propagation}

The framework of PAL can easily be extended to propagate Gauss-Newton targets.
As shown in Ref.~\cite{meulemans2020theoretical}, this requires the training of feedback connections such that they invert the signal passed from given layer $\ell$ to the output layer $N$;
i.e.~training weights such that the backward mapping Jacobian matrix $\bm J_{g_{\ell,N}}$ is a pseudoinverse of the forward mapping, $\bm J_{g_{\ell,N}} = \big[\bm J_{f_{N,\ell}} \big]^+$.
These feedback mappings could be realized by skip connections from the output layer to each hidden layer, while maintaining a layer-wise feed-forward architecture.
In contrast to the layer-wise feedback setup defined in the main text (cf.~Fig.~\ref{fig:mc_setup}), only one interneuron population matching the output layer pyramidal cells would be required here, as the same error signal $\bm e_N$ is backpropagated to all hidden layers.

In analogy to DTP-DRL~\cite{meulemans2020theoretical}, we can define a difference-based reconstruction loss,
\begin{align}
    \mathcal{L}^\mathrm{diff}_\ell = \|\bm B_{\ell,N} \, \bm{\widehat{r}}_N - \bm \xi_\ell\|^2 + \alpha \, \|\bm B_\ell \|^2 \;.
\end{align}
Gradient descent on this reconstruction loss yields the update rule
\begin{align}
    \bm \dot{\bm{B}}_{\ell, N} &=  -\eta^\text{bw}_{i}  \big[\big(\bm B_{\ell,N} \, \bm{\widehat{r}}_N  - \bm \xi_\ell \big) \, \bm{\widehat{r}}_N + \alpha \, \bm B_{\ell, N}  \big] \;.
\end{align}
If noise injection and plasticity of top-down weights is phased, i.e.~by a schedule that sequentially injects noise only into a given layer $\ell$ while enabling plasticity of $\bm B_{\ell, N}$, this learning rule minimizes the reconstruction loss of $\bm \xi_\ell$ as it passes to the output layer and back.
This can be seen by plugging in the equivalent of Eq.~\eqref{eq:app_rP_noise_expanded} for phased noise,
\begin{align}
	\bm{\widehat{r}}_N \approx \varphi'(\bm \ubreven_{N}) \, \big[ \prod_{n=\ell}^{N-1} \bm W_{n+1,n} \varphi'(\bm \ubreven_n) \big] \,\bm \xi_\ell \;.
\end{align}
The difference $\bm B_{\ell,N} \, \bm{\widehat{r}}_N - \bm \xi_\ell$ is minimal if $\bm B_{\ell,N} \, \varphi'(\bm \ubreven_{N})$ is equal to $\lambdaP \, \big[ \prod_{n=\ell}^{N-1}   \bm \WPP_{n+1,n} \varphi'(\bm \ubreven_n) \big]^+$,
thereby aligning the backwards Jacobian with the Moore-Penrose inverse of the forward Jacobian matrix.

As shown in Ref.~\cite{meulemans2020theoretical}, learning backwards weight to minimize such a reconstruction loss produces forward updates closely related to Gauss-Newton optimization,
\begin{align}
    \Delta \bm \WPP_{\ell,\ell-1} \propto \big[{\bm J}_{f_{N,\ell}}\big]^+ \bm e_N \, (\bm \rP_{\ell-1})^T \;,
\end{align}
where ${\bm J}_{f_{N,\ell}}$ denotes the Jacobian matrix mapping potentials in layer $\ell$ to the output layer $N$.

While error propagation using the Moore-Penrose inverse has been shown to perform well on simple classification tasks~\cite{lee2015difference,bartunov2018assessing,meulemans2020theoretical,NEURIPS2020_72619259}, it is currently not known how such a difference reconstruction loss could be implemented while training all top-down weights simultaneously.

\section{Simulation of PAL}

\begin{algorithm}
\DontPrintSemicolon
\SetKwInOut{Parameters}{Parameters}
\KwIn{Data stream encoded as rate vector $\bm \rP_0[t]$, target voltage vector $\bm u^\text{tgt}[t] $}
\Parameters{Network with layers 1 to $N$; effective neuron time constants $\taueffP_\ell, \taueffI_\ell$}
\SetKwBlock{Begin}{Update instantaneous rates and compartment potentials}{end function}
\Begin()
{
  \For {$\ell$ in $range(1,N)$}
  {
    $\bm {\breve u^\text{P}}_\ell[t] \gets \bm { u^\text{P}}_\ell[t-dt] + \frac{\taueffP_\ell}{dt} \; (\bm { u^\text{P}}_\ell[t] - \bm { u^\text{P}}_\ell[t-dt])  $\\
    $\bm {\breve u}^\text{I}[t] \gets \bm { u^\text{I}}[t-dt] + \frac{\taueffI_\ell}{dt} \; (\bm { u^\text{I}}[t] - \bm { u^\text{I}}[t-dt])  $\\
    ${\bm r}^\text{P}_\ell[t] \gets \varphi\big(\bm {\breve u^\text{P}}_\ell[t]\big)$\\
    ${\bm r}^\text{I}_\ell[t] \gets \varphi\big(\bm {\breve u}^\text{I}_\ell[t]\big)$
  }
  \For {$\ell$ in $range(1,N)$}
  {
    $\bm v^\bas_\ell[t] \gets \bm \WPP_{\ell, \ell-1}[t] \; \bm \rP_{\ell-1} [t] $ \\
    $\bm v^\api_\ell[t] \gets \bm \BPP_{\ell, \ell+1}[t] \; \bm \rP_{\ell+1} [t] + \bm \BPI_{\ell, \ell}[t] \; \bm \rI_\ell [t] $\\
    $\bm v^\den_\ell[t] \gets \bm \WIP_{\ell, \ell}[t] \; \bm r^\text{P}_{\ell} [t] $
  }
}

\SetKwBlock{Begin}{Update high-pass filtered rates}{end function}
\Begin()
{
\For {$\ell$ in $range(1,N)$}
    {
$\widehat{\bm r}^\text{P}_\ell[t] \gets \widehat{\bm r}^\text{P}_{\ell}[t-dt] + {\bm r}^\text{P}_{\ell}[t] - {\bm r}^\text{P}_{\ell}[t-dt] - \frac{dt}{\tauhp} \widehat{\bm r}^\text{P}_{\ell}[t-dt]$
    }

}

\SetKwBlock{Begin}{Update noise vectors}{end function}
\Begin()
{
  \For {$\ell$ in $range(1,N-1)$}
  {
    $ \bm \mu_\ell \sim \mathcal N (0, \mathbb{I})$\\
    $\bm \xi_\ell[t] \gets \bm \xi_\ell[t-dt] + \frac{1}{\tauxi} \big( \sqrt{\tauxi \, dt} \, \sigma_\ell \,  \bm \mu_\ell -  dt \; \bm \xi_\ell[t-dt] \big)$
  }
}

\SetKwBlock{Begin}{Update somatic potentials with noise injection}{end function}
\Begin()
{
  \For {$\ell$ in $range(1,N-1)$}
  {
    $\bm u_\ell^\text{eff,I}[t] \gets  \taueffI_\ell \, \big\{g^\den \,\bm v_\ell^\den[t] + g^\text{nudge,I}\, \bm \ubreveP_{\ell+1}[t]\big\} $\\
    $\Delta \bm u_\ell^\text{I}[t] \gets \frac{dt}{\taueffI_\ell} \; \big\{\bm u_\ell^\text{eff,I}[t] - \bm u_\ell^\text{I}[t] \big\} $\\
    $\bm u^\text{eff,P}_\ell[t] \gets  \taueffP_\ell \, \big\{g^\bas \,\bm v^\bas_\ell[t] + g^\api\, (\bm v^\api_\ell[t]+\bm \xi_\ell[t])\big\} $\\
    $\Delta \bm u^\text{P}_\ell[t] \gets \frac{dt}{\taueffP_\ell} \; \big\{\bm u^\text{eff,P}_\ell[t] - \bm u^\tP_\ell[t] \big\} $
  }
    $\bm u^\text{eff,P}_N[t] \gets  \taueffP_N \, \big\{g^\bas \,\bm v^\bas_N[t] + g^\text{nudge,tgt}\, \bm u^\text{tgt}[t]\big\} $\\
    $\Delta \bm u^\text{P}_N[t] \gets \frac{dt}{\taueffP_N} \; \big\{\bm u^\text{eff,P}_N[t] - \bm u^\tP_N[t] \big\} $\\
      apply voltage updates: $\bm u^\text{P}_\ell + \Delta \bm u^\text{P}_\ell, \bm u^\text{I}_\ell + \Delta \bm u^\text{I}_\ell\;\; \forall\, \ell$
}

\SetKwBlock{Begin}{Update weights, incl.~low-pass filtering feedforward  weight updates}{end function}
\Begin()
{
  \For {$\ell$ in $range(1,N-1)$}
  {
        $\Delta \bm \WPP_{\ell, \ell-1} \gets dt \; \eta_\ell^\mathrm{fw} \big\{\bm { \rP}_\ell[t] - \varphi \big(\frac{g^\bas}{\gl + g^\bas + g^\api} \bm v^\bas_\ell[t-dt] \big) \big\}  \; \bm \rP_{\ell-1}[t-dt]$\\
        $\Delta \bm \BPP_{\ell, \ell+1} \gets dt \;\eta^\text{bw}_{\ell}  \big\{ \bm \xi_\ell[t]\, \widehat{\bm r}^\text{P}_{\ell+1}[t-dt] - \alpha_\ell \, \bm \BPP_{\ell, \ell+1} [t] \big\}$\\
        $\Delta \bm \BPI_{\ell, \ell} \gets - dt \; \eta^\text{PI}_{\ell} \, \bm v^\api_\ell[t-dt]\, \bm \rI_\ell[t-dt]$\\
    $\Delta \bm \WIP_{\ell, \ell} \gets dt \; \eta^\text{IP} \big\{ \bm \rI_\ell[t] - \varphi\big(\frac{g^\den}{\gl + g^\den} \bm v^\den_\ell[t-dt] \big) \big\} \;  \bm {{r}}^\text{P}_{\ell}[t-dt]$
    }
    $\Delta \bm \WPP_{N, N-1} \gets dt \; \eta_\ell^\mathrm{fw} \big\{\bm { \rP}_N[t] - \varphi \big(\frac{g^\bas}{\gl + g^\bas } \bm v^\bas_N[t-dt] \big) \big\}  \; \bm \rP_{N-1}[t-dt]$\\
  \For {$\ell$ in $range(1,N)$}
  {
    $\overline{\Delta \bm W}^\text{PP}_{\ell, \ell-1} \gets \overline{\Delta \bm W}^\text{PP}_{\ell, \ell-1}[t-dt] + \frac{dt}{\taulo} \big( \Delta \bm \WPP_{\ell, \ell-1}[t-dt] - \overline{\Delta \bm W}^\text{PP}_{N, N-1}[t-dt] \big)$\\
    }
    apply weight updates: $\bm \WPP_{\ell,\ell-1} + \overline{\Delta \bm W}^\text{PP}_{\ell,\ell-1}, \bm \BPP_{\ell,\ell+1} + \Delta \bm \BPP_{\ell,\ell+1}, \bm \BPI_{\ell,\ell} + \Delta \bm \BPI_{\ell,\ell}, \bm \WIP_{\ell,\ell} + \Delta \bm \WIP_{\ell,\ell}\;\; \forall\, \ell$
}

\caption{Dendritic cortical microcircuits with prospective coding and PAL}\label{algo:PAL}
\end{algorithm}

\label{app:sims}

\subsection{Microcircuit models}

\begin{table}[h]
\centering
\caption{Parameters for microcircuit model simulations.}
\begin{threeparttable}
\begin{tabular}{cccccc}
 	& Fig.~\ref{fig:bw_learning} (a+b+c)  & Fig.~\ref{fig:bw_learning} (d+e+f) & Fig.~\ref{fig:teacher_student} (a+b) & Fig.~\ref{fig:teacher_student} (c+d) & Fig.~\ref{fig:teacher_student} (e+f) \\ \hline
$dt$~[ms]	& $10^{-2}$  & $10^{-2}$ & $10^{-2}$ & $10^{-2}$ & $10^{-2}$ \\
$\Tpres$~[ms]	& 1  & 1 & 1 & 1 & 1 \\
$\tauhp$~[ms]	& 0.1  & 0.1  & 0.1 & 0.1 & 0.1 \\
$\taulo$~[ms]	& ---\tnote{$\dagger$}  & ---\tnote{$\dagger$} & $10^2$ & $10^2$ & $10^2$ \\
$\tauxi$~[ms]	& 0.1  & 0.1 & 0.1 & 0.1 & 0.1 \\
noise scale $\sigma_\ell\, \forall \, \ell$ & $5 \times 10^{-2} $  & $5 \times 10^{-2}$ & $ 10^{-2} $ &  $ 10^{-2} $ & $ 10^{-2} $ \\
regularizer $\alpha$ & $10^{-5}$  & $10^{-5}$ & $10^{-6}$ & $10^{-6}$ & $10^{-6}$ \\ \hline
$g_l$~[ms$^{-1}$]	&0.03&0.03 &0.03 & 0.03 & 0.03\\
$g^\text{bas}$~[ms$^{-1}$]	&0.1&0.1 &0.1 & 0.1 & 0.1\\
$g^\text{api}$~[ms$^{-1}$]	&0.06&0.06 &0.06 & 0.06 & 0.06\\
$g^\text{den}$~[ms$^{-1}$]	&0.1&0.1 &0.1 & 0.1 & 0.1\\
$g^\text{nudge, I}$~[ms$^{-1}$]	&0.06&0.06 &0.06 & 0.06 & 0.06\\
$g^\text{nudge,tgt}$~[ms$^{-1}$]	&0.06&0.06 &0.06 & 0.06 & 0.06\\ \hline
input & $\mathcal{U}[0, 1]$ & $\mathcal{U}[0, 1]$  & $\mathcal{U}[0, 1]$  & Yin-Yang & MNIST\\
dataset size & 100 & 100 & 100 & 6\,000 & 50\,000 \\
epochs & 100 & 500 & 5\,000 & 400 & 100 \\
network size & [5-20-10-20-5] & [5-20-10-20-5]  & [1-1-1] & [4-30-3] & [784-100-10]\\
activation & $\frac{1}{1+e^{-x}}$ & $\frac{1}{1+e^{-x}}$ & $\frac{1}{1+e^{-x}}$ & $\frac{1}{1+e^{-x}}$ & $\frac{1}{1+e^{-x}}$ \\ \hline
$\eta^\text{fw}$~[ms$^{-1}$] & $0, 0, 0, 0$ & $0, 0, 0, 0$ & $2,0.5$  & $50, 0.01$ & $1.0, 5\times 10^{-3}$ \\
$\eta^\text{bw}$~[ms$^{-1}$] & $50, 50, 50$  & $20, 20, 20$ & 20 & 0.5 & 0.2 \\
$\eta^\text{IP}$~[ms$^{-1}$] & $0, 0, 0$ & $0, 0, 0$ & 10 & 0.05 & 0.02 \\
$\eta^\text{PI}$~[ms$^{-1}$] & $5, 5, 5$ & $0.5, 0.5, 0.5$ & 0.5  & 0.02 & 0.02\\ \hline
weight init $\bm \WPP_{\ell, \ell-1}$ & $\mathcal{U}[-1, 1]$ & $\mathcal{U}[-5, 5]$ & $\mathcal{U}[-1, 0]$ & $\mathcal{U}[-0.1, 0.1]$ & $\mathcal{U}[-0.1, 0.1]$\\
weight init $\bm \BPP_{\ell-1, \ell}$ & $\mathcal{U}[-1, 1]$ & $\mathcal{U}[-5, 5]$ & $\mathcal{U}[-1, 0]$ & $\mathcal{U}[-1, 1]$ &$\mathcal{U}[-1, 1]$\\
\end{tabular}
\vspace{0.3cm}
\begin{tablenotes}
\footnotesize
\item[$\dagger$] No forward weight updates applied.
\end{tablenotes}
\end{threeparttable}
\label{tab:sim_values_mc}
\end{table}

Below we address several implementation details:

For all simulations, we set the resting potential to $\bm \El =0$.
\added{Effective time constants are calculated as the ratio of dimensionless conductances (setting the capacitance to 1):
	$\taueffP_\ell =  \frac{1}{\gl + \gbas + \gapi}$ for pyramidal neurons in hidden layers;
	$\taueffP_N =  \frac{1}{\gl + \gbas + g^\text{nudge,tgt}}$ for output layer neurons in presence of a nudging signal; 
	$\taueffP_N =  \frac{1}{\gl + \gbas}$ for output layer neurons in absence of a nudging signal; $\taueffI_\ell =  \frac{1}{\gl + \gden + g^\text{nudge,I}}$ for interneurons.}

Fig.~\ref{fig:bw_learning} (Phaseless backwards weight alignment):
In order to compare the backprojections with those in an ANN trained with BP (right column), we perform several steps.
After each epoch of top-down weight training, we instantiate a teacher model with the architecture as the respective `student' microcircuit model.
To this model, we pass the input sequence and record the teacher output as a target signal.
We now provide the newly acquired input/target pairs to the student model.
As in all other simulations, these pairs are presented for $\Tpres = 100 \, dt$, and we disable noise during this evaluation.
It is also necessary to set the lateral weights to the self-predicting state defined by Eqs.~\eqref{eq:app_the_SPS} in order to obtain a measurable error signal in layers below the last hidden layer.
We record the potential weight updates defined by the backprojections, but do not apply them.

Next, we feed the same input/target sequence into an ANN with weights set to $\bm \WPP_{\ell, \ell-1}$ and layer size and activation as defined in Tab.~\ref{tab:sim_values_mc}.
For this ANN, we calculate the output layer error as the difference between target and voltage (linear activation on output layer).
The weight update given in this ANN is then compared to the recorded update for the microcircuit model.

Fig.~\ref{fig:teacher_student} (Teacher-student setup):
We initialize the teacher with weights $\bm \WPP_{2, 1} = \bm \WPP_{1, 0} = 2$.
For the student models with feedback alignment, the same parameters as for PAL are used, but with $\eta^\text{bw} = \eta^\text{PI} = 0$, no noise injection, and without the low-pass filter on $\Delta \bm \WPP_{\ell, \ell-1}$.
For the BP reference model, we additionally set $\bm \BPP_{\ell-1, \ell} = [\bm \WPP_{\ell, \ell-1}]^T$ and $\bm \BPI_{\ell-1, \ell} = - \bm \BPP_{\ell-1, \ell}$ after every update.

Fig.~\ref{fig:teacher_student} (Classification tasks):
For the Yin-Yang task, we used datasets of size 6\,000 for training, 900 for validation, and 900 for testing.
For MNIST digit classification, sets were 50\,000 for training, 10\,000 for validation, and 10\,000 for testing.
In either case, FA runs use the same parameters as PAL, but with fixed $\bm \BPP_{\ell, \ell+1}$ and $\bm \BPI_{\ell, \ell}$ and no noise injection.
The reference network is an ANN trained with BP using a Cross-Entropy loss and ADAM with default parameters of PyTorch.
In this case, 10 seeds were trained.

\clearpage

\subsection{Efficient credit assignment in deep networks}

We detail the parameters for the general LI-model simulation below:
\begin{table}[h]
	\centering
	\caption{Parameters for deep learning simulations of PAL}
\begin{threeparttable}
	\begin{tabular}{ccc}
		& Fig.~\ref{fig:eff_deep_nets} & Fig.~\ref{fig:cifar_10}  \\ \hline
		$dt$~[ms]	& $10^{-1}$& $10^{-1}$  \\
		$\Tpres$~[ms]	& 10 & 10  \\
		$\tauhp$~[ms]	& 1& 1  \\
		$\taulo$~[ms]	& ---\tnote{$\dagger$}& ---\tnote{$\ddagger$}\\
		$\tauxi$~[ms]	& 1& 1 \\
		noise scale $\sigma_\ell $ & $10^{-2} \, \forall \, \ell$ & $5 \cdot 10^{-2} \, \forall \, \ell$   \\
		regularizer $\alpha$ & $10^{-4}$ & $5\times 10^{-5}$ \\ \hline
		input & MNIST & CIFAR-10 \\
		loss & MSE & MSE \\
		epochs & 10 & 50 \\
		batch size & 32 & 128 \\
		architecture & Dense:~[784-200-2-200-784] & Conv2d($(5\times 5) \times 20$), MaxPool(2)\\
		 & & Conv2d($(5\times 5) \times 50$), MaxPool(2) \\
		 &  & 500 \\
		 &  & 10 \\
		activation & $\tanh$, linear, $\tanh$, linear & $\text{sigmoid}$, $\text{sigmoid}$, $\text{sigmoid}$, linear \\ \hline
		$\eta^\text{fw}$~[ms$^{-1}$] & $8 \times 10^{-3}, 8 \times 10^{-3}, 8 \times 10^{-3}, 8 \times 10^{-4}$  & $5, 1, 2, 0.2$  \\
		$\eta^\text{bw}$~[ms$^{-1}$] & $0.8, 0.8, 0.8$ & $0\tnote{$\star$} \;, 10, 10$ \\ \hline
		weight init $\bm W_{\ell, \ell-1}$ & $\mathcal{N}(0, 0.05)$ & $\mathcal{N}(0, 0.05)$ \\
		weight init $\bm B_{\ell-1, \ell}$ & $\mathcal{N}(0, 0.05)$ &  $\mathcal{N}(0, 0.05)$ \\
		bias init $\bm b_\ell$ & $\mathcal{N}(0, 0.05)$ & $\mathcal{N}(0, 0.05)$ \\
	\end{tabular}
\begin{tablenotes}
\footnotesize
\item[$\dagger$]Low-pass filter disabled, as we found no improvement when enabled.
\item[$\ddagger$]Noise does not affect top-down errors in this setup (see Methods); thus, low-pass filtering of weights is disabled.
\item[$\star$]PAL enabled only for fully connected layers, see Methods.
\end{tablenotes}
\end{threeparttable}
\label{tab:sim_values_DL}
\end{table}

Errors on the output layer are defined as $\bm e_N = \beta \, (\bm u^\text{tgt} -  \bm \ubreve_N)$, and we have simulated with $\beta = 0.1$.
The dataset sizes for the MNIST autoencoder task are 50\,000 for training, 10\,000 for validation, and 10\,000 for testing.
For CIFAR-10, this was 40\,000 training images, 10\,000 for validation, and 10\,000 for testing.

\clearpage

\subsection{PAL outperforms FA with noise}

\begin{figure}[h]
	\centering
	\includegraphics[width=.7\textwidth]{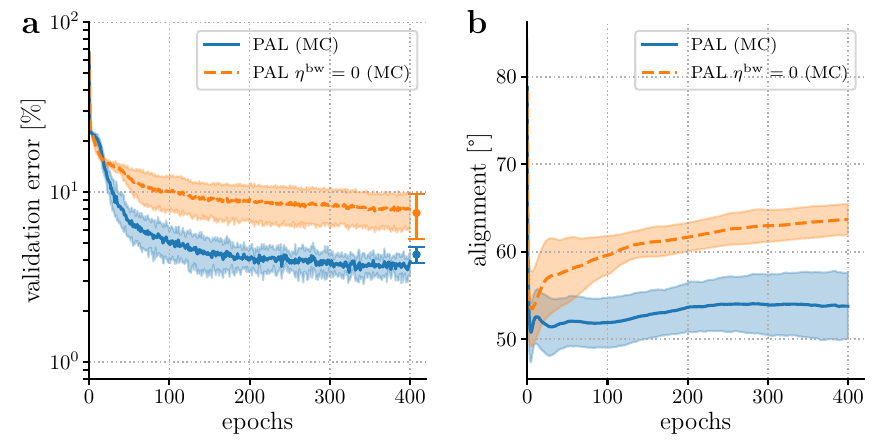}
	\caption{
		\added{\textbf{PAL outperforms FA not simply due to inclusion of noise.}
			A key difference between the experiments performed with PAL and FA is the inherent modeling of noise.
			Therefore, it could be argued that FA with noise may perform on par with PAL.
			To test this, we reproduce the Yin-Yang experiment of Fig.~\ref{fig:teacher_student} (c,d).
			PAL without learning of top-down weights ($\eta^\text{bw}=0$) is equivalent to FA with noise, which performs similar to vanilla FA and is still outperformed by PAL.
		}
	}
	\label{fig:FA_with_noise}
\end{figure}

\newpage

\subsection{PAL makes full use of prospective coding for efficient weight learning}

\begin{figure}[h]
	\centering
	\includegraphics[width=.5\textwidth]{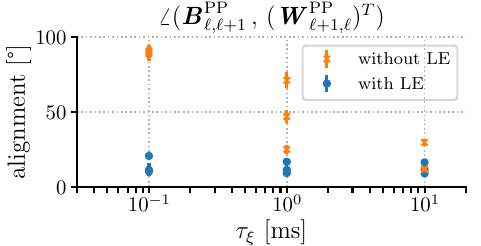}
	\caption{
		\added{
			\textbf{PAL makes full use of prospective coding for efficient weight learning.}
			The setup of Fig.~\ref{fig:bw_learning}~(a,b,c) is repeated (learning of top-down weights in the microcircuit model)
			with and without prospective coding (Latent Equilibrium, LE).
			Shown are the angles between forward and backward weights for all layers after 100 epochs (mean and standard deviation over 10~seeds).
			In order to learn the correct weights, the model without prospective coding needs to be trained with orders of magnitude larger time constants 
			(the noise time constant $\tauxi$ needs to be larger than $\taueffP \approx 7$~ms, while $\Tpres \gg \tauxi$ still holds).
			Prospective coding therefore represents a leap in efficiency for learning in cortical models with slow neurons.
			Note that this advantage also translates to practical terms -- in this case, wall clock simulation times increased from 10~minutes to 15 hours when disabling prospective coding.
		}
	}
	\label{fig:PAL_needs_LE}
\end{figure}

\end{document}